\documentclass[a4paper,11pt]{article}

\usepackage{mydebug}
\usepackage[utf8x]{inputenc}
\usepackage{latexsym}
\usepackage{amsmath}
\usepackage{verbatim}
\usepackage{enumerate}
\usepackage{amssymb}

\usepackage{csquotes}
\usepackage{graphicx}
\usepackage{textcomp}
\usepackage[ruled,vlined,linesnumbered]{algorithm2e}
\usepackage{epsfig}
\usepackage{amsfonts}
\usepackage{todonotes}
\usepackage[inline]{enumitem}
\usepackage{longtable}
\usepackage{xspace}
\usepackage{url}
\usepackage{subcaption}

\usepackage[colorlinks = true, backref=page]{hyperref}
\usepackage{amsthm}
\usepackage{amsmath}
\usepackage{amssymb}
\usepackage{amsfonts}
\usepackage{amstext}

\usepackage{xspace}
\usepackage{graphicx}
\usepackage[mathscr]{euscript}
\usepackage{fullpage}	

\usepackage{color, colortbl}
\definecolor{LRed}{rgb}{1,.8,.8}
\definecolor{MRed}{rgb}{1,.6,.6}
\definecolor{HRed}{rgb}{1,.2,.2}

\hypersetup{colorlinks={true},linkcolor={red},citecolor=blue}

\usepackage[noabbrev,capitalise]{cleveref}

\usetikzlibrary{decorations.markings}

\tikzset{->-/.style={decoration={
  markings,
  mark=at position .5 with {\arrow{>}}},postaction={decorate}}}

\newcommand{\ioctfull}{\textsc{Stable Odd Cycle Transversal}}
\newcommand{\ioct}{\textsc{SOCT}}
\newcommand{\istsfull}{{\sc Stable} $s$-$t$ {\sc Separator}}
\newcommand{\ists}{\textsc{SSTS}}
\newcommand{\idfvsfull}{\textsc{Stable Directed Feedback Vertex Set}}
\newcommand{\idfvs}{\textsc{SDFVS}}

\newcommand{\aoctfull}{\textsc{Annotated Odd Cycle Transversal}}
\newcommand{\aoct}{\textsc{AOCT}}
\newcommand{\astsfull}{{\sc Annotated} $s$-$t$ {\sc Separator}}
\newcommand{\asts}{\textsc{ASTS}}

\newcommand{\octfull}{\textsc{Odd Cycle Transversal}}
\newcommand{\oct}{\textsc{OCT}}

\newcommand{\dfvsfull}{\textsc{Directed Feedback Vertex Set}}
\newcommand{\dfvs}{\textsc{DFVS}}

\newcommand{\stablebip}{\textsc{Stable Bipartization}}

\newcommand{\tw}{\mathbf{tw}}

\newcommand{\dg}{{\sf degeneracy}}

\DeclareMathAlphabet{\w}{OT1}{pzc}{m}{it}

\newcommand{\cO}{\mathcal{O}}

\newcommand{\F}{{\mathcal F}}
\newcommand{\OO}{{\mathcal O}}
\newcommand{\yes}{{\sf Yes}}
\newcommand{\no}{{\sf No}}

\newcommand{\FPT} {{\sf FPT}\xspace}

\newcommand{\NPC} {{\sf NP}-complete\xspace}
\newcommand{\WOH} {{\sf W}$[1]$-hard\xspace}

\newtheorem{theorem}{Theorem}
\newtheorem{lemma}{Lemma}[section]
\newtheorem{claim}{Claim}[section]
\newtheorem{corollary}{Corollary}
\newtheorem{definition}{Definition}[section]
\newtheorem{observation}{Observation}[section]
\newtheorem{proposition}{Proposition}[section]

\newtheorem{fact}{Fact}[section]

\crefname{claim}{claim}{claims}
\crefname{fact}{fact}{facts}

\newcommand{\h}[1]{\end{document}}

\usepackage{boxedminipage}
 
\newcommand{\eps}{\varepsilon}

\usepackage{bold-extra}
\usepackage{lmodern}
\usepackage[T1]{fontenc}
\usepackage{textcomp}

\newcommand{\ifam}[2]{\ensuremath{{\mathscr F}(#1,#2)}}
\newcommand{\I}{\cal{I}}

 \newcommand{\defparproblem}[4]{
\noindent\fbox{
  \begin{minipage}{0.96\textwidth}
   \begin{tabular*}{\textwidth}{@{\extracolsep{\fill}}lr} #1  & {\bf{Parameter:}} #3
 \\ \end{tabular*}
  {\bf{Input:}} #2  \\
  {\bf{Question:}} #4
  \end{minipage}
  }
 \vspace{1mm}
}

\newcommand{\dpc}[2]{$#1$-$#2$-digraph pair cut}

\newcommand{\sep}[2]{$#1$-$#2$-separator}
\newcommand{\imsep}[2]{important $#1$-$#2$-separator} 
\newcommand{\imseps}[2]{important $#1$-$#2$-separators}
\newcommand{\imc}{\textsc{Stable Multicut}}

\newcommand{\SSS}{{\mathcal S}}
\newcommand{\mc}{\textsc{Multicut}}

\title{Covering Small Independent Sets and Separators with Applications to Parameterized Algorithms\thanks{Supported by Pareto-Optimal Parameterized Algorithms, ERC Starting Grant 715744,  Parameterized Approximation, ERC Starting Grant 306992, 
and  Rigorous Theory of Preprocessing, ERC Advanced Investigator Grant 267959.}}
\author{{\large Daniel Lokshtanov\thanks{University of Bergen, Norway. \texttt{\{daniello|fahad.panolan|meirav.zehavi\}@ii.uib.no}}}\addtocounter{footnote}{-1}
\and{\large Fahad Panolan}\footnotemark \addtocounter{footnote}{-1}
\and {\large Saket Saurabh\footnotemark~\thanks{Institute of Mathematical Sciences, India. \texttt{\{saket|roohani\}@imsc.res.in}}}\addtocounter{footnote}{-1}
\and{\large Roohani Sharma}\footnotemark\addtocounter{footnote}{-2}
\and{\large  Meirav Zehavi}\footnotemark
}

\date{}
\begin{document}
 \maketitle

\thispagestyle{empty}
\begin{abstract}
We present two new combinatorial tools for the design of parameterized algorithms. The first is a simple linear time randomized algorithm that given as input a $d$-degenerate graph $G$ and an integer $k$, outputs an independent set $Y$, such that for every independent set $X$ in $G$ of size at most $k$, the probability that $X$ is a subset of $Y$ is at least $\left({(d+1)k \choose k} \cdot k(d+1)\right)^{-1}$.
%
%
%
The second is a new (deterministic) polynomial time graph sparsification procedure that given a graph $G$, a set $T = \{\{s_1, t_1\}, \{s_2, t_2\}, \ldots, \{s_\ell, t_\ell\}\}$ of terminal pairs and an integer $k$, returns an induced subgraph $G^\star$ of $G$ that maintains {\em all} the inclusion minimal multicuts of $G$ of size at most $k$, and does not contain any $(k+2)$-vertex connected set of size $2^{\OO(k)}$. In particular, $G^\star$ excludes a clique of size $2^{\OO(k)}$ as a topological minor.
Put together, our new tools yield new randomized fixed parameter tractable (\FPT) algorithms for {\sc Stable} $s$-$t$ {\sc Separator}, {\sc Stable Odd Cycle Transversal} and {\sc Stable Multicut} on general graphs, and for {\sc Stable Directed Feedback Vertex Set} on $d$-degenerate graphs, resolving two problems left open by Marx et al. [ACM Transactions on Algorithms, 2013]. All of our algorithms can be derandomized at the cost of a small overhead in the running time.

\end{abstract}

\newpage
\setcounter{page}{1}

\newcommand{\kptc}{64^{k+2}  \cdot (k + 2)^2}
\newcommand{\Zsize}{16^{k+1} \cdot 64(k+2)}
\newcommand{\Wsize}{64^{k+2} \cdot 4(k + 2)^2}

\section{Introduction}
We present two new combinatorial tools for designing parameterized algorithms. The first is a simple linear time randomized algorithm that given as input a $d$-degenerate graph $G$ and an integer $k$, outputs an independent set $Y$, such that for every independent set $X$ in $G$ of size at most $k$, the probability that $X$ is a subset of $Y$ is at least $\left({k(d+1) \choose k} \cdot k(d+1)\right)^{-1}$. Here, an {\em independent set} in a graph $G$ is a vertex set $X$ such that no two vertices in $X$ are connected by an edge, and the {\em degeneracy} of an $n$-vertex graph $G$ is the minimum integer $d$ such that there exists an ordering $\sigma : V(G) \rightarrow \{1, \ldots, n\}$ such that every vertex $v$ has at most $d$ neighbors $u$ with $\sigma(u) > \sigma(v)$. Such an ordering $\sigma$ is called a $d$-{\em degeneracy sequence} of $G$. We say that a graph is $d$-degenerate, if $G$ has a $d$-degeneracy sequence. More concretely, we prove the following result.


\begin{lemma}\label{lem:indcover}
There exists a linear time randomized algorithm that given as input a $d$-degenerate graph $G$ and an integer $k$, outputs an independent set $Y$, such that for every independent set $X$ in $G$ of size at most $k$ the probability that $X$ is a subset of $Y$ is at least $\left({k(d+1) \choose k} \cdot k(d+1)\right)^{-1}$. 
\end{lemma}

\begin{proof}
Given $G$, $k$ and a $d$-degeneracy sequence $\sigma$ of $G$ the algorithm sets $p = \frac{1}{d+1}$ and colors the vertices of $G$ black or white independently with the following probability : a vertex gets color black with probability $p$ and white with probability $1-p$. The algorithm then constructs the set $Y$ which contains every vertex $v$ that is colored black and all the neighbors $u$ of $v$ with $\sigma(u) > \sigma(v)$ are colored white. 
We first show that $Y$ is an independent set. Suppose not. Let $u,v \in Y$, such that $\sigma(u) < \sigma(v)$ and $uv \in E(G)$. Since $u \in Y$, by the construction of $Y$, $v$ has to be colored white. This contradicts that $v \in Y$ because every vertex in $Y$ is colored black.

We now give a lower bound on the probability with which a given independent set $X$ of size at most $k$ is contained in $Y$.
Define $Z$ to be the set of vertices $u$ such that $u$ has a neighbor $x \in X$ with $\sigma(x) < \sigma(u)$. Since every $x \in X$ has at most $d$ neighbors $u$ with $\sigma(x) < \sigma(u)$, it follows that $|Z| \leq kd$.
Observe that $X \subseteq Y$ precisely when all the vertices in $X$ are colored black and all the vertices in $Z$ are colored white. This happens with probability 
$$p^{|X|}(1-p)^{|Z|} \geq \left(\frac{k}{k(d+1)}\right)^k \cdot \left(\frac{kd}{k(d+1)}\right)^{kd} \geq \left[{(d+1)k \choose k} \cdot k(d+1)\right]^{-1}.$$
Here, the last inequality follows from the fact that binomial distributions are centered around their expectation. This concludes the proof.
\end{proof}

\Cref{lem:indcover} allows us to reduce many problems with an independence constraint to the same problem without the independence requirement. For an example, consider the following four well-studied problems: 
\begin{itemize}\setlength\itemsep{-.5mm}
\item {\sc Minimum} $s$-$t$ {\sc Separator}: Here, the input is a graph $G$, an integer $k$ and two vertices $s$ and $t$, and the task is to find a set $S$ of at most $k$ vertices such that $s$ and $t$ are in distinct connected components of $G - S$. This is a classic problem solvable in polynomial time~\cite{ford1956maximal,stoer1997simple}.
\item {\sc Odd Cycle Transversal}: Here, the input is a graph $G$ and an integer $k$, and the task is to find a set $S$ of at most $k$ vertices such that $G - S$ is bipartite. This problem is \NPC~\cite{ChoiNR89} and has numerous fixed-parameter tractable (\FPT) algorithms~\cite{ReedSV04,LokshtanovNRRS14}. For all our purposes, the $\OO(4^k \cdot k^{\OO(1)} \cdot (n+m))$ time algorithms of Iwata et al.~\cite{IwataOY14} and Ramanujan and Saurabh~\cite{ramanujan2014linear} are the most relevant. 
\item {\sc Multicut}: Here, the input is a graph $G$, a set $T = \{\{s_1,t_1\}, \{s_2, t_2\}, \ldots,  \{s_\ell, t_\ell\}\}$ of terminal pairs and an integer $k$, and the task is to find a set $S$ on at most $k$ vertices such that for every $i \leq \ell$, $s_i$ and $t_i$ are in distinct connected components of $G - S$. Such a set $S$ is called a {\em multicut} of $T$ in $G$. This problem is \NPC even for $3$ terminal pairs, that is, when $l=3$~\cite{edgemultiwaycuthardness}, but it is \FPT~\cite{BousquetDT11,MarxR14} parameterized by $k$, admitting an algorithm~\cite{LokshtanovRS16} with running time $2^{\OO(k^3)} \cdot mn\log n$.
\item {\sc Directed Feedback Vertex Set}: Here, the input is a directed graph $D$ and an integer $k$, and the task is to find a set $S$ on at most $k$ vertices such that $D - S$ is acyclic. This problem is also \NPC~\cite{karp1972reducibility} and \FPT~\cite{chen2008fixed} parameterized by $k$, admitting an algorithm~\cite{LokshtanovRS16} with running time $\OO(k!\cdot 4^k \cdot k^5 \cdot (n + m))$.
\end{itemize}

In the \enquote{stable} versions of all of the above-mentioned problems, the solution set $S$ is required to be an independent set\footnote{Independent sets are sometimes called stable sets in the literature. In this paper we stick to independent sets, except for problem names, which are inherited from Marx et al.~\cite{marx2013finding}.}. Fernau~\cite{DemGMS07} posed as an open problem whether {\sc Stable Odd Cycle Transversal} is \FPT. This problem was resolved by Marx et al.~\cite{marx2013finding}, who gave \FPT algorithms for {\sc Stable} $s$-$t$ {\sc Separator} running in time $2^{2^{k^{O(1)}}} \cdot (n+m)$ and {\sc Stable Odd Cycle Transversal} running in time $2^{2^{k^{\OO(1)}}} \cdot (n+m) + \OO(3^k \cdot nm)$. Here, the $\OO(3^k \cdot nm)$ term in the runnning time comes from a direct invocation of the algorithm of Reed et al.~\cite{ReedSV04} for {\sc Odd Cycle Transversal}. Furthermore, Marx et al.~\cite{marx2013finding} gave an algorithm for {\sc Stable Multicut} with running time $f(k, |T|)(n+m)$ for some function $f$. They posed as open problems, the probelem of determining whether there exists an \FPT algorithm for {\sc Stable Multicut} parameterized by $k$ only, and the problem of determining whether there exists an \FPT algorithm for {\sc Stable Odd Cycle Transversal} with running time $2^{k^{O(1)}}\cdot (n+m)$.

Subsequently, algorithms for {\sc Odd Cycle Transversal} with running time $4^kk^{\OO(1)}\cdot (n+m)$ were found, independently by Iwata et al.~\cite{IwataOY14} and Ramanujan and Saurabh~\cite{ramanujan2014linear}. Replacing the call to the algorithm of Reed et al.~\cite{ReedSV04} in the algorithm of Marx et al.~\cite{marx2013finding} for {\sc Stable Odd Cycle Transversal} by any of the two $4^k \cdot k^{\OO(1)} \cdot (n+m)$ time algorithms for {\sc Odd Cycle Transversal} yields a  $2^{2^{k^{O(1)}}} \cdot (n+m)$ time algorithm for {\sc Stable Odd Cycle Transversal}. However, obtaining a  $2^{k^{O(1)}}(n+m)$ time algorithm still a remained open problem.

Using \Cref{lem:indcover}, we directly obtain randomized \FPT algorithms for {\sc Stable} $s$-$t$ {\sc Separator}, {\sc Stable Odd Cycle Transversal}, {\sc Stable Multicut} and {\sc Stable Directed Feedback Vertex Set} on $d$-degenerate graphs. It is sufficient to apply \Cref{lem:indcover} to obtain an independent set $Y$ containing the solution $S$, and then run the algorithms for the non-stable version of the problem where all vertices in $V(G) \setminus Y$ are not allowed to go into the solution. For all of the above-mentioned problems, the existing algorithms can easily be made to work even in the setting where some vertices are not allowed to go into the solution.

\Cref{lem:indcover} only applies to graphs of bounded degeneracy. Even though the class of graphs of bounded degeneracy is quire rich (it includes planar graphs, and more generally all graphs excluding a topological minor) it is natural to ask whether \Cref{lem:indcover} could be generalized to work for all graphs. However, if $G$ consists of $k$ disjoint cliques of size $n/k$ each, the best success probability one can hope for is $(k/n)^k$, which is too low to be useful for \FPT algorithms. 

At a glance the applicability of \Cref{lem:indcover} seems to be limited to problems on graphs of bounded degeneracy. However, there already exist powerful tools in the literature to reduce certain problems on general input graphs to special classes. For us, the {\em treewidth reduction} of Marx et al.~\cite{marx2013finding} is particularly relevant, since a direct application of their main theorem reduces {\sc Stable} $s$-$t$ {\sc Separator} and {\sc Stable Odd Cycle Transversal} to the same problems on graphs of bounded treewidth. Since graphs of bounded treewidth have bounded degeneracy, we may now apply our algorithms for bounded degeneracy graphs, obtaining new \FPT algorithms for {\sc Stable} $s$-$t$ {\sc Separator} and {\sc Stable Odd Cycle Transversal} on general graphs. Our algorithms have running time $2^{k^{\OO(1)}} \cdot (n+m)$,
thus resolving, in the affirmative, one of the open problems of Marx et al.~\cite{marx2013finding}. 

One of the reasons that the parameterized complexity of {\sc Stable Multicut} parameterized by the solution size was left open by  Marx et al.~\cite{marx2013finding} was that their treewidth reduction does not apply to multi-terminal cut problems when the number of terminals is unbounded. Our second main contribution is a graph sparsification procedure that works for such multi-terminal cut problems. Given a graph $G$ and a set $T$ of terminal pairs, a multicut $S$ of $T$ in $G$ is called a {\em minimal} multicut o $T$ in $G$ if no proper subset of $S$ is a multicut of $T$ in $G$. A vertex set $X$ in $G$ is {\em vertex-$k$-connected} (or just $k$-{\em connected})  if, for every pair $u$, $v$ of vertices in $X$, there are $k$ internally vertex disjoint paths from $u$ to $v$ in $G$.

\begin{theorem}\label{thm:multicutReduction}
There exists a polynomial time algorithm that given a graph $G$, a set $T = \{\{s_1, t_1\}, \{s_2, t_2\}, \ldots, \{s_\ell, t_\ell\}\}$ of terminal pairs and an integer $k$, returns an induced subgraph $G^\star$ of $G$ and a subset $T^\star$ of $T$ which have the following properties.
\begin{itemize}\setlength\itemsep{-.5mm}
\item every minimal multicut of $T$ in $G$ of size at most $k$ is a minimal multicut of $T^*$ in $G^\star$,
\item every minimal multicut of $T^*$ in $G^\star$ of size at most $k$ is a minimal multicut of $T$ in $G$, and
\item $G^\star$ does not contain a $(k+2)$-connected set of size $\OO(64^k \cdot k^2)$.
\end{itemize}
\end{theorem}

We remark that excluding a $(k+2)$-connected set of size $\OO(64^k \cdot k^2)$ implies that $G^\star$ excludes a clique of size $\OO(64^k \cdot k^2)$ as a {\em topological minor}. In fact, the property of excluding a large $(k+2)$-connected set puts considerable extra restrictions on the graph, on top of being topological minor free, as there exist planar graphs that contain arbitrarily large $(k+2)$-connected sets. The proof of \Cref{thm:multicutReduction} uses the {\em irrelevant vertex} technique of Robertson and Seymour~\cite{RobertsonS95b}, however, instead of topological arguments for finding an irrelevant vertex we rely on a careful case distinction based on cut-flow duality together with counting arguments based on {\em important separators}.

\Cref{thm:multicutReduction} reduces the {\sc Stable Multicut} problem on general graphs to graphs excluding a clique of size $2^{\OO(k)}$ as a topological minor. Since such graphs have bounded degeneracy~\cite{BollobasT98,KomloS96}, our algorithm for {\sc Stable Multicut} on graphs  of bounded degeneracy yields an \FPT algorithm for the problem on general graphs, resolving the second open problem posed by Marx et al.~\cite{marx2013finding}.

We remark that a sparsification for directed graphs similar to \Cref{thm:multicutReduction} powerful enough to handle {\sc Directed Feedback Vertex Set} is unlikely, since {\sc Stable Directed Feedback Vertex Set} on general graphs is known to be \WOH~\cite{misra2012parameterized}, while our algorithm works on digraphs where the underlying undirected graph has bounded degeneracy.

The algorithms based on \Cref{lem:indcover} are randomized, however they can be derandomized using a new combinatorial object that we call $k$-independence covering families, that may be of independent interest. We call an independent set of size at most $k$ a $k$-{\em independent set}, and we call a family of independent sets an {\em independent family}.  An independent family $\mathscr{F}$ {\em covers} all $k$-independent sets of $G$, if for every $k$-independent set $X$ in $G$ there exists an independent set $Y \in \mathscr{F}$ such that $X \subseteq Y$. In this case, we call ${\mathscr F}$ a $k$-{\em independence covering family}. An algorithm based on \Cref{lem:indcover} can be made deterministic by first constructing a $k$-independence covering family $\mathscr{F}$, and then looping over all sets $Y \in \mathscr{F}$ instead of repeatedly drawing $Y$ at random using \Cref{lem:indcover}.

Since a graph $G$ contains at most $n^k$ independent sets of size at most $k$, drawing $\OO({k(d+1) \choose k} \cdot kd \cdot \log n)$ sets using \Cref{lem:indcover} and inserting them into $\mathscr{F}$ will result in a $k$-independence covering family with probability at least $1/2$. Hence, for every $d$ and $k$, every graph $G$ on $n$ vertices of degeneracy at most $d$ has a $k$-independence covering family of size at most $\OO({k(d+1) \choose k} \cdot kd \cdot \log n)$. By direct applications of existing pseudo-random constructions (of {\em lopsided universal sets}~\cite{FominLPS16} and {\em perfect hash families}~\cite{FredmanKS84}) we show that given a graph $G$ of degeneracy $d$ and integer $k$ one can construct a $k$-independence covering family of size not much larger than $\OO({k(d+1) \choose k} \cdot kd \cdot \log n)$ in time roughly proportionate to its size. 

Additionally, we give an efficient construction of $k$-independence covering families of size at most $f(k)n$ for {\em nowhere dense} class of graphs~\cite{nevsetvril2008grad,nevsetvril2011nowhere}, based on low treedepth colorings of such graphs. This construction immediately yields \FPT algorithms for the considered problems on nowhere dense classes of graphs.

\subsection{Proof sketch for \Cref{thm:multicutReduction}.}
Towards the proof of Theorem~\ref{thm:multicutReduction}, we describe an algorithm that given $G$, the set $T$ of terminal pairs, an integer $k$ and a $(k+2)$-connected set $W$ of size at least $\kptc$, computes a vertex $v$ that does not appear in any minimal multicut of size at most $k+1$.  One can show that such a vertex $v$ is {\em irrelevant} in the sense that $G$, $T$ has {\em exactly} the same family of minimal multicuts of size at most $k$ as the graph $G - v$ with the terminal set $T' = \{\{s_i, t_i\} \in T : v \notin \{s_i, t_i\}\}$. The proof of Theorem~\ref{thm:multicutReduction} then follows by repeatedly removing irrelevant vertices, until $|W| \leq \kptc$.

\smallskip
\noindent
{\bf Degree $1$ Terminals Assumption.}
In order to identify an irrelevant vertex it is helpful to assume that every terminal $s_i$ or $t_i$ has degree $1$ in $G$, and that no vertex in $G$ appears in more than one terminal pair. To justify this assumption one can, for every pair $\{s_i, t_i\} \in T$, add $k+2$ new degree $1$ vertices $s_i^1, s_i^2, \ldots, s_i^{k+2}$ and make them adjacent to $s_i$, and $k+2$ new degree 
$1$ vertices $t_i^1, t_i^2, \ldots, t_i^{k+2}$ and make them adjacent to $t_i$. Call the resulting graph $G'$, and make a terminal pair set $T'$ from $T$ by inserting for every pair $\{s_i, t_i\} \in T$ the set $\{\{s_i^j, t_i^j\} : 1 \leq j \leq k+2\}$ into $T'$. It is clear that the set of (minimal) multicuts of $T'$ in $G'$ of size at most $k+1$ is the same as the set of (minimal) multicuts of $T$ in $G$ of size at most $k+1$.

\smallskip
\noindent
{\bf Detecting Irrelevant Vertices.}
In order to identify an irrelevant vertex we investigate the properties of all vertices $v \in W$ for which there exists a minimal multicut of size at most $k+1$ containing $v$. We will call such vertices {\em relevant}. Let $v \in W$ be a relevant vertex and let $S$ be a minimal multicut of size at most $k+1$ containing $v$. Since $W$ is a $(k+2)$-connected set and $|S| \leq k+1$, $W \setminus S$ is contained in some connected component $C$ of $G - S$. Since $S$ is a multicut we also have that $S$ is a {\em pair cut} for $T$ with respect to $W$ in the following sense: for each terminal pair $\{s_i, t_i\}$ at most one of $s_i$ and $t_i$ can reach $W \setminus S$ in $G - S$. This is true because all vertices of $W \setminus S$ lies in the same connected component of $G - S$. Furthermore $S \setminus \{v\}$ can {\em not} be a pair cut for $T$ with respect to $W$, because if it happened to be a pair cut, then we can show that $S \setminus \{v\}$ would also have been a multicut, contradicting the minimality of $S$. We say that $v \in W$ is {\em essential} if there exists some pair cut $S$ for $T$ with respect to $W$ such that $|S| \leq k+1$, $v \in S$ and $S \setminus \{v\}$ is not a pair cut for $T$ with respect to $W$. The above argument shows that every relevant vertex is essential, and it remains to find a vertex $v \in W$ which is provably not essential.

The algorithm that searches for a non-essential vertex $v$ crucially exploits {\em important separators}, defined by Marx~\cite{Marx06i}. Given a graph $G$ and two vertex sets $A$ and $B$, an $A$-$B$-{\em separator} is a vertex set $S \subseteq V(G)$ such that there is no path from $A \setminus S$ to $B \setminus S$ in $G - S$. An $A$-$B$-separator $S$ is called a {\em minimal} $A$-$B$-{\em separator} if no proper subset of $S$ is also an $A$-$B$-separator. Given a vertex set $S$, we define the {\em reach of $A$ in $G - S$} as the set $R_G(A,S)$ of vertices reachable from $A$ by a path in $G - S$. We can now define a partial order on the set of minimal $A$-$B$ separators as follows. Given two minimal $A$-$B$ separators $S_1$ and $S_2$, we say that $S_1$ is ``at least as good as''  $S_2$ if $|S_1| \leq |S_2|$ and $R_G(A, S_2) \subset R_G(A, S_1)$. In plain words, $S_1$ \enquote{costs less} than  $S_2$ in terms of the number of vertices deleted and $S_1$ \enquote{is pushed further towards $B$} than $S_2$ is. A minimal $A$-$B$-separator $S$ is an {\em important} $A$-$B$-{\em separator} if no minimal $A$-$B$-separator other than $S$ is at least as good as $S$. A key insight behind many parameterized algorithms~\cite{chen2008fixed,Chitnis:2012DSFVS,ChitnisHM13,CyganPPW13,Kratsch:2012MCDAG,LokshtanovM13,LokshtanovR12,LokshtanovRS15,LokshtanovRSULC16,MarxR14} is that for every $k$, the number of important $A$-$B$-separators of size at most $k$ is at most $4^k$~\cite{ChenLL09}. We refer the reader to Marx~\cite{Marx06i} and the textbook by Cygan et al.~\cite{CyganFKLMPPS15} for a more thorough exposition of important separators.

The algorithm that searches for a non-essential vertex $v$  makes the following case distinction. Either there exists a small $T$-$W$-separator $Z$, or there are many vertex disjoint paths from $T$ to $W$. 
Here ,we have abused notation by treating $T$ as a set of vertices in the terminal pairs rather than a set of terminal pairs. 
In the first case, when there exists a $T$-$W$-separator $Z$ of size at most $\zeta =\Zsize$, we show that every  relevant vertex $v \in W$ is contained in some important $z$-$W$-separator of size at most $k+1$, for some $z\in Z$. Since there are at most $4^{k+1}$ such important separators and we can enumerate them efficiently~\cite{ChenLL09}, the algorithm simply marks all the vertices in $W$ appearing in such an important separator and outputs one vertex that is not marked.

\smallskip
\noindent
{\bf Many Disjoint Paths.}
If there are at least $\Zsize$ vertex disjoint paths from $T$ to $W$ we identify a terminal pair $\{s_i, t_i\}$ such that, for every minimal multicut $S$ of size at most $k+1$ for the instance $G$ with terminal set $T \setminus \{\{s_i, t_i\}\}$, $S$ is also a minimal multicut for $G$ with terminal set $T$. Such a terminal pair is irrelevant in the sense that removing $\{s_i, t_i\}$ from $T$ does not change the family of minimal multicuts of size at most $k+1$. Thus, if we later identify a vertex $v \in W$ that is irrelevant with the reduced terminal set, then $v$ is also irrelevant with respect to the original terminal set. We will say that a terminal pair that isn't irrelevant is {\em relevant}.

To identify an irrelevant terminal pair we proceed as follows. Without loss of generality, there are $\zeta/2$ vertex disjoint paths from $A = \{s_1, s_2, \ldots s_{\zeta/2}\}$ to $W$. Thus, for any set $S$ of at most $k+2$ vertices, all of $A$ except for at most $k+2$ vertices can reach $W \setminus S$ in $G - S$. Let $B = \{t_1, t_2, \ldots t_{\zeta/2}\}$. We have that for every pair cut $S$ for $T$ with respect to $W$, at most $k+2$ vertices of $B \setminus S$ are reachable from $W$ in $G - S$. 

Consider a pair $\{s_i, t_i\}$ with $s_i \in A$ and $t_i \in B$. If $\{s_i, t_i\}$ is relevant, then there must exist a set $S$ of size at most $k+1$ that is a minimal pair cut for $G$ with terminals $T \setminus \{\{s_i, t_i\}\}$ with respect to $W$, but is not a pair cut with terminal pair set $T$. We have that $t_i$ is reachable from $W$ in $G-S$, and that $S \cup \{t_i\}$ is a pair cut for $T$. Let $\hat{B} \subseteq B$ be the set of vertices in $B$ that are reachable from $W$ in $G - (S \cup \{t_i\})$. From the discussion in the previous paragraph it follows that $|\hat{B}| \leq k+2$. Thus, $S \cup \{t_i\} \cup \hat{B}$ is a $W$-$B$ separator of size at most $2(k+2)$. Pick any minimal $W$-$B$ separator $\hat{S} \subseteq S \cup \{t_i\} \cup \hat{B}$. 

We argue that $t_i \in \hat{S}$. To that end we show that there exists a path $P$ from $W$ to $t_i$ in $G - (S \cup \hat{B})$. Thus, if $t_i \notin \hat{S}$ then  $\hat{S}$ would be a subset of $S \cup \hat{B}$ and $P$ would be a path from $W$ to $B$ in $G - \hat{S}$, contradicting that $\hat{S}$ is a $W$-$B$-separator. We know that there exists a path $P$ from $W$ to $t_i$ in $G - S$ and that $P$ does not visit any vertex in $\hat{B}$ on the way to $t_i$ because all vertices in $\hat{B}$ have degree $1$. Hence $P$ is disjoint from $\hat{S}$, yielding the desired contradiction. We conclude that $t_i \in \hat{S}$. 

With all of this hard work we have, under the assumption that $\{s_i, t_i\}$ is a relevant pair with $t_i \in B$, exhibited a minimal $W$-$B$-separator $\hat{S}$ that contains $t_i$. There must exist some important $W$-$B$-separator $S^\star$ that is at least as good as $\hat{S}$.  Since all the vertices of $P$ (except $t_i$) are reachable from $W$ in $G - \hat{S}$ it follows that $t_i \in S^\star$. We have now shown that if $\{s_i, t_i\}$ is a relevant pair with $t_i \in B$, then there exists a $W$-$B$ important separator of size at most $2(k+2)$ that contains $t_i$. The algorithm goes over all $W$-$B$ important separators of size at most $2(k+2)$ and marks all vertices appearing in such important separators. Since $\zeta/2 > 4^{2(k+2)} \cdot 2(k+2)$ it follows that some vertex $t_i$ in $B$ is left unmarked. The pair $(s_i, t_i)$ is then an irrelevant pair. This concludes the proof sketch that there exists a polynomial time algorithm that given $G$, $T$, $k$ and $W$ finds an irrelevant vertex in $W$, provided that $W$ is large enough.

\smallskip
\noindent
{\bf Finding a Large $(k+2)$-Connected Set.} We have shown how to identify an irrelevant vertex given a $(k+2)$-connected set $W$ of large size. But how to find such a set $W$, if it exists? Given $G$ we can in polynomial time build an auxiliary graph $G^*$ that has the same vertex set as $G$. Two vertices in $G^*$ are adjacent if there are at least $k+2$ internally vertex disjoint paths between them in $G$. Clearly $k+2$-connected sets in $G$ are cliques in $G^*$ and vice versa. However, finding cliques in general graphs is $W[1]$-hard, and believed to not even be approximable in FPT time. To get around this obstacle we exploit the special structure of $G^*$.

A $(k+2)$-connected set $W$ in $G$ of size at least $\Wsize$ induces a subgraph of $G^*$ where every vertex has degree at least $(k+2)$. Thus the degeneracy of $G^*$ is at least $\Wsize$. A modification of a classic result of Mader~\cite{Mader1972} (see also Diestel~\cite{diestel} and lecture notes of Sudakov~\cite{Sudakov16}) shows that every graph of degeneracy at least $4d$ contains a $(d+1)$-connected set of size at least $d+2$, and that such a set can be computed in polynomial time. We apply this result with $d = \kptc-1$ to obtain a $(\kptc)$-connected set in $W^*$ in $G^*$ of size at least $\kptc$. A simple argument shows that $W^*$ is also a $(k+2)$-connected set in $G$. We may now apply the algorithm to detect irrelevant vertices using $W^*$. This concludes the proof sketch of Theorem~\ref{thm:multicutReduction}.

\smallskip
\noindent
{\bf Guide to the paper.} In \Cref{sec:prelims} we introduce basic notations and some well known results needed for our work. 
In \Cref{sec:ISCLemma} we define independence covering families and give constructions of such families. This allows to derandomize algorithms based on \Cref{lem:indcover}.
%
We then construct independence covering families for nowhere dense classes of graphs, and show some barriers to further generalizations of our results. A reader content with randomized \FPT algorithms may skip this section altogether. 
%
%
%
In \Cref{sec:AppI} we show the applicability of \Cref{lem:indcover} (or independence covering families) by designing \FPT algorithms for {\sc Stable} $s$-$t$ {\sc Separator}, {\sc Stable Odd Cycle Transversal}, {\sc Stable Multicut} and {\sc Stable Directed Feedback Vertex Set} on $d$-degenerate graphs. 
%
In \Cref{sec:AppII}, we explain how the algorithms from \Cref{sec:AppI} combined with the treewidth reduction procedure of Marx et al.~\cite{marx2013finding} lead to \FPT algorithms for some of the considered problems on general graphs.
%
In \Cref{sec:thmone} we prove \Cref{thm:multicutReduction}. This is the most technically challenging part of the paper, and may be read independently of the other sections. 


\section{Preliminaries}\label{sec:prelims}

 


We use ${\mathbb N}$ to denote the set of natural numbers starting from $0$. For 
$t\in {\mathbb N}$, $[t]$ is a shorthand  for $\{1,\ldots,n\}$. 
The symbol $\log$ denotes natural logarithm and $e$ denotes the base of 
natural logarithm.  For a set $U$ and $t\in {\mathbb N}$, we use $2^U$ and $\binom{U}{t}$ to denote the power set of $U$ and the set of subsets of $U$ of size $t$, respectively. 
For a function $f : D \to R$, $X \subseteq D$ and $Y\subseteq R$, we denote $f(X) =\{ f(x) :  x\in X\}$ 
and $f^{-1}(Y) = \{d :  f(d)\in Y \}$.

\begin{fact}
\label{fact:binom}
$\frac{1}{n}\left[{\left(\frac{k}{n}\right)}^{{-k}} {\left(\frac{n-k}{n}\right)}^{{-(n-k)}}\right] \leq \binom{n}{k}
\leq \left[{\left(\frac{k}{n}\right)}^{{-k}} {\left(\frac{n-k}{n}\right)}^{-(n-k)}\right]$.
\end{fact}

\paragraph*{Graphs}
Throughout our presentation, given a (di)graph $G$, $n$ denotes the number of vertices in $G$ and $m$ denotes the number of (arcs)edges in $G$.
We use the term {\em graphs} to represent undirected graphs.  
For a (di)graph $G$, $V(G)$ denotes its vertex set, $A(G)$ denotes arc set in case of digraphs, and  $E(G)$ denotes edge set 
in case of graphs. For any positive integers $a,b$, we denote by $K_{a,b}$ the complete bipartite graph with $a$ vertices in one part and $b$ vertices in the other part.
Let $G$ be a (di)graph.  
For any $X \subseteq V(G)$, $G[X]$ denotes the induced graph on the vertex set $X$.
By $G- X$, we denote the (di)graph $G[V(G)\setminus X]$. 
When $X=\{v\}$, we use $G-v$ to denote the graph $G-\{v\}$. 
For a set $Y \subseteq E(G)$, $G-Y$ denotes the (di)graph obtained from $G$ by deleting the edges in $Y$. For any $u,v \in V(G)$, $d_G(u,v)$ denotes the number of (arcs)edges on the shortest path from $u$ to $v$ in $G$.
For a graph $G$, for any $u,v \in V(G)$, $uv$ denotes the edge with endpoints $u$ to $v$. For any $v \in V(G)$, $N_G(v)$ denotes the neighbors of $v$ in $G$, that is, $N_G(v)=\{u: uv\in E(G)\}$. The degree of a vertex $v$ in $G$, denoted by $deg_G(v)$, is equal to the number of neighbors of $v$ in $G$, that is, $deg_G(v) = \vert N_G(v) \vert$. The minimum degree of $G$ is the minimum over the degrees of all its vertices. If $D$ is a digraph, then for any $u,v \in V(D)$, $uv$ denotes the arc from $u$ to $v$.
By $\overleftarrow{D}$, we denote the digraph obtained from $D$ by reversing each of its arcs. For any $v \in V(D)$, $N^{+}_D(v)$ denotes the out-neighbors of $v$ in $D$ and $N^{-}_D(v)$ denotes the in-neighbors of $v$ in $D$, that is, $N^{+}_D(v) =\{u : vu \in A(D)\}$ and $N^{-}_D(v) = \{u : uv \in A(D)\}$. For any $X \subset V(D)$, $N^{+}_D(X) = \{ u : u \in V(D) \setminus X \text{ and there exists } v \in X \text{ such that } vu \in A(D)\}$ and $N^{-}_D(X) =\{u : u \in V(D) \setminus X \text{ and there exists } v \in X \text{ such that } uv \in A(D)\}$.
For a graph $G$, $\tw(G)$ denotes the treewidth of $G$. 

For a non-negative integer $d$, a graph $G$ is called a \emph{$d$-degenerate graph} if for every subgraph $H$ of $G$ there exists $v \in H$ such that $deg_H(v) \leq d$. The degeneracy of a graph $G$, denoted by $\dg(G)$, is the least integer $d$, for which $G$ is $d$-degenerate. 
If there exists a subgraph $H$ of $G$ such that the minimum degree of $H$ is at least $d$, we say that the degenaracy of $G$ is at least $d$. For a $d$-degenerate graph $G$, a \emph{$d$-degenaracy sequence} of $G$ is an ordering of the vertices of $G$, say $\sigma : V(G) \to [|V(G)|]$, such that $\sigma$ is a bijection and, for any $v \in V(G)$, $|N_G(v) \cap \{u : \sigma(u) > \sigma(v)\}| \leq d$. For a given degenaracy sequence $\sigma$ and a vertex $v \in V(G)$, the vertices in $N_G(v) \cap \{u : \sigma(u) > \sigma(v)\}$ are called the \emph{forward neighbors} of $v$ in $\sigma$, and this set of forward neighbors is denoted by $N_{G,\sigma}^f(v)$. The following proposition says 
we can find $d$-degeneracy sequence of a graph in linear time.  

\begin{proposition}[\cite{matula1983smallest}]\label{prop:deg}
If $G$ is a $d$-degenerate graph, for some non-negative integer $d$, then a $d$-degenaracy sequence of $G$ exists and can be found in time $\OO(n+m)$.
\end{proposition}

%


\paragraph*{Graphs Separators} 
For (di)graph $G$, $X,Y \subseteq V(G)$, an \emph{\sep{X}{Y}} in $G$ is a subset $C\subseteq V(G)$, such that 
there is no path from a vertex in $X\setminus C$ to a vertex in $Y\setminus C$ in $G - C$.
For (di)graph $G$, $s,t \in V(G)$ an \emph{\sep{s}{t}} in $G$ is a subset $C\subseteq V(G)\setminus \{s,t\}$ 
such that there is no path from $s$ to $t$ in $G - C$. Similarly, for $X,Y \subseteq V(G)$, a $(X,Y)$ separator is a subset $C \subseteq V(G)$ such that there is no path from any vertex of $X$ to any vertex of $Y$ in $G - C$.
The size of a separator is equal to the cardinality of the separator. A \emph{minimum \sep{s}{t}} in $G$ is the one with the minimum number of vertices. 
A set $Y \subseteq V(G)$ is a \emph{min vertex cut (mincut)} of $G$ if $G - Y$ has at least two components.




Since, checking whether there is an \sep{s}{t} of weight at most $k$ (here there is an integer weight function on $V(G)$ is given) can be done by running at most $k$ rounds of the classical Ford-Fullkerson algorithm, \Cref{prop:ststime} follows.

\begin{proposition}\label{prop:ststime}
Given a (di)graph $G$, $s,t \in V(G)$, an integer $k$ and $w: V(G) \to \mathbb{N}$, an \sep{s}{t} of weight at most $k$, if it exists, can be found in time $\OO(k \cdot (n+m))$. Also, a minimum \sep{s}{t} can be found in time $\OO(mn)$. 
\end{proposition}

\begin{proposition} [\cite{stoer1997simple}] \label{prop:mincuttime}
A mincut of a (di)graph $G$ can be found in time $\OO(m + n \log n)$.
\end{proposition}

%
%

\newcommand{\succprob}{\frac{1}{\binom{k(d+1)}{k} \cdot (k(d+1))}}
\newcommand{\Fcard}{(e(d+1))^{k} \cdot 2k^2(d+1) \cdot \log n}
\newcommand{\minus}{-}
\newcommand{\icf}{{independence covering family}\xspace}

\section{Tool I: Independence Covering Lemma}\label{sec:ISCLemma}
In this section we give constructions of   $k$-independence covering families, which are useful in derandomizing algorithms based on \Cref{lem:indcover}.  Towards this we first formally define the notion of  $k$-independence covering family -- a family of independent sets of a graph $G$ which  covers all independent sets in $G$ of size at most $k$. 

\begin{definition}[$k$-Independence Covering Family]
For a graph $G$ and $k\in {\mathbb N}$, a family  of independent sets of $G$ is called an {\icf} for $(G,k)$, denoted by $ \ifam{G}{k}$, if for any independent set $X$ in $G$ of size at most $k$, there exists $Y \in  \ifam{G}{k}$ such that $X \subseteq Y$.  
\end{definition}

 Observe that for any pair $(G,k)$, there exists an \icf of size at most $\binom{n}{k}$ containing all independent 
sets of size at most $k$. We show that, if $G$ has bounded degeneracy, then $k$-\icf of \enquote{small} size exists.  In fact, we give both randomized and deterministic algorithms to construct such a family of \enquote{small} size for graphs of  bounded degeneracy.  
In particular, we prove that if $G$ is $d$-degenerate, then one can construct an independent set covering family for $(G,k)$ of size $f(k,d) \cdot \log n$, where $f$ is a function depending only on $k$ and $d$.  We first give the randomized algorithm for constructing $k$-independence covering family. Towards this we use the algorithm described in \Cref{lem:indcover}. 
For an ease of reference we present the algorithm given in  \Cref{lem:indcover} here.

\begin{algorithm}[h]
Construct a $d$-degeneracy sequence $\sigma$ of $G$, using \Cref{prop:deg}.\label{step:d:degcon}\\
Set $p = \frac{1}{d+1}$. Independently color each vertex $v\in V(G)$ black with probability $p$ and 
white with probability $(1-p)$. \label{step1:randproc} \\
Let $B$ and $W$ be the set of vertices colored black and white, respectively. \\
$Z:=\{v\in B~|~N_{G,\sigma}^f(v)\cap B=\emptyset\}$. \label{stepZ}\\ 
\Return{Z}
\caption{Input is $(G,k)$, where $G$ is a $d$-degenerate graph and $k\in {\mathbb N}$}
\label{alg:onerandomprocess}
\end{algorithm}

\begin{lemma}[Randomized Independence Covering Lemma]\label{lemma:riscl} 
There is an algorithm that given a $d$-degenerate graph $G$ and $k\in{\mathbb N}$, 
  outputs a family $ \ifam{G}{k}$ such that $(a)$ 
$ \ifam{G}{k}$ is an independence covering family for $(G,k)$ with probability at least $1-\frac{1}{n}$, 
$(b)$ $\vert  \ifam{G}{k} \vert \leq \binom{k(d+1)}{k} \cdot 2k^2(d+1) \cdot \log n$, and $(c)$ the running time of the algorithm is $\OO(\vert  \ifam{G}{k} \vert \cdot (n+m))$.  
\end{lemma}
\begin{proof}

Let $t=\binom{k(d+1)}{k} \cdot k(d+1)$. 
We now explain the algorithm to 
construct the family $ \ifam{G}{k}$  mentioned in the lemma. We run \Cref{alg:onerandomprocess} (\Cref{lem:indcover})  
$\gamma=t\cdot 2k\log n$ times. Let $Z_1,\ldots , Z_{\gamma}$ be the sets that are output at the end of each iteration of \Cref{alg:onerandomprocess}. Let $\ifam{G}{k} = \{Z_1,\ldots Z_{t\cdot 2k\log n}\}$. Clearly,  
$\vert \ifam{G}{k} \vert = t\cdot 2k \log n  =  \binom{k(d+1)}{k} \cdot 2k^2(d+1) \cdot \log n$. 
Thus condition $(b)$ is proved. 
The running time of the algorithm (condition $(c)$) follows from \Cref{lem:indcover}. 

Now we prove condition $(a)$ of the lemma. 
Fix an independent set $X$ in $G$ of cardinality at most $k$. 
By \Cref{lem:indcover}, we know that for any $Z\in \ifam{G}{k}$, 
$\Pr[X\subseteq Z]\geq \frac{1}{t}$.  Thus the probability that there does not exist 
a set $Z\in \ifam{G}{k}$ such that $X\subseteq Z$ is at most $(1-\frac{1}{t})^{\vert \ifam{G}{k} \vert}\leq e^{-2k \log n}=n^{-2k}$. The last inequality follows from a well-known fact that $(1-a)\leq e^{-a}$ for any $a\geq 0$. Since the total number of independent sets of size at most $k$ in $G$ 
is upper bounded by $n^{k}$, by the union bound, the probability that there exists an independent set of size at most $k$ which is not a subset of any set in $\ifam{G}{k}$ is upper bounded by $n^{-2k} \cdot n^{k} = n^{-k} \leq 1/n$. 
This implies that $\ifam{G}{k}$ is  an \icf for $(G,k)$ with probability at least $1-\frac{1}{n}$. 
\end{proof}

Together with \Cref{fact:binom}, \Cref{lemma:riscl} implies the following corollary.

\begin{corollary}
The cardinality of the family $\ifam{G}{k} $ constructed by the algorithm of \Cref{lemma:riscl} is at most $\Fcard$.
\end{corollary}
\begin{proof}
From \Cref{lemma:riscl},
\begin{align*}
\vert \ifam{G}{k}  \vert & =  \binom{k(d+1)}{k} \cdot 2k^2(d+1) \cdot \log n \\
&\leq (d+1)^{k} \cdot \left(\frac{1+d}{d}\right)^{kd} \cdot 2k^2(d+1) \cdot \log n \qquad\qquad (\mbox{By \Cref{fact:binom}})\\
&= \left[(d+1) \left(1+\frac{1}{d}\right)^{d} \right]^{k} \cdot 2k^2(d+1) \cdot \log n\\
& \leq (e(d+1))^{k} \cdot 2k^2(d+1) \cdot \log n \qquad\qquad (\mbox{Because} \left(1+\frac{1}{d}\right)^{d}\leq e, \forall d >0). 
\end{align*}
This completes the proof. 
\end{proof}

\paragraph*{Deterministic Constructions.} 
The two deterministic algorithms, that we give, are obtained from the randomized algorithm presented in \Cref{lemma:riscl} by using the $n$-$p$-$q$-lopsided-universal family~\cite{FominLPS16} and the $(n,\ell,\ell^2)$-perfect hash family~\cite{FredmanKS84}, respectively. 
Both of our deterministic constructions basically replace the random coloring of the vertices in \Cref{step1:randproc} of \Cref{alg:onerandomprocess} by a coloring defined by $n$-$p$-$q$-lopsided universal family and the $(n,\ell,\ell^2)$-perfect hash family, respectively. 
In the following, we first define the $n$-$p$-$q$-lopsided-universal family, state \Cref{prop:lopsided} (an algorithm to construct an $n$-$p$-$q$-lopsided family of) which is followed by our first deterministic algorithm (\Cref{lemma:discl}).

\begin{definition}[$n$-$p$-$q$-lopsided-universal family~\cite{FominLPS16}]
A family $\mathcal{F}$ of sets over a universe $U$ is an $n$-$p$-$q$-lopsided-universal family if for every $A \in {U \choose p}$ and $B \in {{U \setminus A} \choose q}$, there is an $F \in \mathcal{F}$ such that $A \subseteq F$ and $B \cap F = \emptyset$.
\end{definition}
\begin{proposition}[Lemma 4.2, \cite{FominLPS16}]\label{prop:lopsided}
There is an algorithm that given $n,p,q\in {\mathbb N}$ and a universe $U$, runs in time $\OO({{p+q} \choose p} \cdot 2^{o(p+q)} \cdot n \log n)$, and outputs an $n$-$p$-$q$-lopsided universal family $\mathcal{F}$ of size at most ${{p+q} \choose p} \cdot 2^{o(p+q)} \cdot \log n$.
\end{proposition}

\begin{lemma}[Deterministic Independence Covering Lemma]\label{lemma:discl} 
There is an algorithm that given a $d$-degenerate graph $G$ and $k\in {\mathbb N}$, 
runs in time $\OO(\binom{k(d+1)}{k} \cdot 2^{o(k(d+1))} \cdot (n+m)\log n)$, 
and outputs a $k$-\icf for $(G,k)$ of size at most $\binom{k(d+1)}{k} \cdot 2^{o(k(d+1))} \cdot \log n$.
\end{lemma}
\begin{proof}
Let $\mathcal{S}$ be the $n$-$p$-$q$-lopsided-universal family constructed using \Cref{prop:lopsided} for $n=|V(G)|$, $p=k$ and $q=kd$. For each $S\in \SSS$, we run \Cref{alg:onerandomprocess}, where \Cref{step1:randproc} is 
replaced as follows: for each vertex in $S$, we color it black and we color all the other vertices white. More precisely, we run \Cref{alg:onerandomprocess} for each $S\in \SSS$, replacing \Cref{step1:randproc} by the procedure just defined, and output the collection $\ifam{G}{k}$ of sets returned at the end of each iteration. 
The size bound on $\vert \ifam{G}{k} \vert$ follows from \Cref{prop:lopsided} and the running time of the algorithm follows from the fact that each run of   \Cref{alg:onerandomprocess} takes $\OO(n+m)$ time.

We now show that $ \ifam{G}{k}$ is, indeed, an independent set covering family for $(G,k)$. Let $X$ be an independent set of cardinality at most $k$ in  $G$. Let $\sigma$ be the $d$-degenerate sequence constructed in \Cref{step:d:degcon} of 
 \Cref{alg:onerandomprocess}. 
Let $Y=\bigcup\limits_{v \in X} N_{G,\sigma}^f(v)$. Since $X$ is independent, $X\cap Y=\emptyset$. 
Furthermore, since $\sigma$ is a $d$-degeneracy sequence and $\vert X\vert\leq k$, we have that $\vert Y \vert \leq kd$. 
By the definition of $n$-$p$-$q$-lopsided-universal family, 
there is a set $S\in\SSS$ such that $X\subseteq S$ and $S\cap Y=\emptyset$. 
Consider the run of \Cref{alg:onerandomprocess} for the set $S$. 
In this run, we have that $X \subseteq B$ and $Y \subseteq W$. 
From the definition of $X,Y$ and $Z$ (set constructed in \Cref{stepZ}), 
we have that $X\subseteq Z$.  This implies that $ \ifam{G}{k}$ is an \icf of $(G,k)$. This completes the proof.  
\end{proof}

Note that when the algorithm of \Cref{lemma:discl} gets as input a graph $G$ whose degeneracy 
is polynomial in $k$, 
the algorithm of \Cref{lemma:discl} returns an \icf $\mathcal{F}$ of size that is single-exponential in $k$.
If the degeneracy of the input graph is exponential in $k$, then the algorithm of \Cref{lemma:discl} can not guarantee a single-exponential sized family. In such cases, our next deterministic algorithm gives better guarantees.
The next deterministic algorithm for computing \enquote{small} sized \icf for  graphs of bounded degeneracy uses the notion of $(n,\ell,\ell^2)$-perfect hash family. The algorithm is described in \Cref{lemma:discl2}. 

\begin{definition}[$(n,\ell,q)$-perfect hash family]
For non-negative integers $n$ and $q$, a family of functions $f_1, \ldots, f_t$ from a universe $U$ of size $n$ to a universe of size $q$ is called a $(n,\ell,q)$-perfect hash family, if for any $S \subseteq U$ of size at most $\ell$, there exists $i \in [t]$, such that $f_i$ is injective on $S$.
\end{definition}

\begin{proposition}[\cite{FredmanKS84}]\label{prop:perfecthash}
For any non-negative integers $n$ and $\ell$, and any universe $U$ on $n$ elements, a $(n,\ell,\ell^2)$-perfect hash family of size $\ell^{\OO(1)} \cdot \log n$ can be computed in time $\ell^{\OO(1)} \cdot n \log n$.  
\end{proposition}

\begin{lemma}\label{lemma:discl2}
There is an algorithm that given a $d$-degenerate graph $G$ and $k\in {\mathbb N}$, 
runs in time $\OO(\binom{k^2 {(d+1)}^2}{k} \cdot {(k(d+1))}^{\OO(1)} \cdot (n+m) \log n)$, and outputs an \icf for $(G,k)$ of size at most $\binom{k^2{(d+1)}^2}{k} \cdot {(k(d+1))}^{\OO(1)} \cdot \log n$.
\end{lemma}
\begin{proof}
Let $\ell=k(d+1)$. Let $\mathcal{S}$ be the $(n,\ell,\ell^2)$-perfect hash family constructed by the algorithm of \Cref{prop:perfecthash}. 
The algorithm replaces the random coloring of vertices in \Cref{step1:randproc} of \Cref{alg:onerandomprocess}, by colorings defined by the $(n,\ell,\ell^2)$-perfect hash family, $\mathcal{S}$. In particular, for each $f \in \mathcal{S}$ and $A \in \binom{[\ell^2]}{k}$, we run \Cref{alg:onerandomprocess}, where \Cref{step1:randproc} is replaced as follows: all vertices in $f^{-1}(A)$ are colored black and the remaining vertices are colored white. More precisely, we run \Cref{alg:onerandomprocess} for each $f\in \SSS$ and $A \in \binom{[\ell^2]}{k}$, replacing \Cref{step1:randproc} by the procedure just defined, and output the collection $\ifam{G}{k}$ of sets returned at the end of each iteration. Clearly, the size of the family constructed at the end of this procedure, is $\vert \mathcal{S} \vert \cdot \binom{\ell^2}{k}$. 
The total running time for this algorithm is $\OO(|\ifam{G}{k}| \cdot (n+m))$. 
Thus, the total running time of this algorithm and the size of the output family as claimed in the lemma follows from \Cref{prop:perfecthash}.

We now show that $\ifam{G}{k}$ is, indeed, an \icf for $(G,k)$. Let $X$ be an independent set of cardinality at most $k$ in $G$. Let $\sigma$ be the $d$-degeneracy sequence constructed in \Cref{step:d:degcon} of  \Cref{alg:onerandomprocess}. 
Let $Y=\bigcup\limits_{v \in X} N_{G,\sigma}^f(v)$. Since $X$ is independent, $X\cap Y=\emptyset$. 
Since $\sigma$ is a $d$-degeneracy sequence and $\vert X\vert\leq k$, we have that $\vert Y \vert \leq kd$. 
Let $P$ and $Q$ be sets of vertices such that, $X \subseteq P, Y \subseteq Q ,|P| = k, |Q| =kd$   
and $P\cap Q=\emptyset$. 
By the definition of $(n,\ell,\ell^2)$-perfect hash family, there exists a function, say $f$, in $\mathcal{S}$, such that $f$ is injective on $P \cup Q$. Now consider the iteration of the algorithm corresponding to $f$ and $f(P)$. 
In this iteration, the algorithm colors all vertices of $P$ with black and all the remaining vertices with white. Since $X \subseteq P$ and $P \cap Q = \emptyset$, the algorithm colors all vertices of $X$ with black color and all vertices of $Y$ with white color. 
From the definitions of $X,Y$ and $Z$ (set constructed in \Cref{stepZ}), 
we have that $X\subseteq Z$. This implies that $\ifam{G}{k}$ is an \icf of $(G,k)$. This concludes the proof. 
\end{proof}

Together with \Cref{fact:binom}, \Cref{lemma:discl2} implies the following corollary.

\begin{corollary}
The size of the family $\ifam{G}{k}$ constructed by the algorithm of \Cref{lemma:discl2} is at most $k^k \cdot {(d+1)}^{2k} \cdot e^{k - \frac{1}{{(d+1)}^2}} \cdot {(k(d+1))}^{\OO(1)} \cdot \log n$.
\end{corollary}
\begin{proof}
Let $\ell = k(d+1)$.
From \Cref{lemma:discl2},
\begin{align*}
\vert \ifam{G}{k} \vert & =  \binom{{\ell}^2}{k} \cdot {\ell}^{\OO(1)} \cdot \log n\\
& \leq {\left(\frac{{\ell}^2}{k}\right)}^{k} \cdot {\left(\frac{{\ell}^2}{{\ell}^2 - k}\right)}^{{\ell}^2 -k} \cdot {\ell}^{\OO(1)} \cdot \log n \qquad\qquad (\mbox{By \Cref{fact:binom}})\\
& = k^k \cdot {(d+1)}^{2k} \cdot {\left ( 1- \frac{k}{{\ell}^2} \right )}^{k- {\ell}^2} \cdot {\ell}^{\OO(1)} \cdot \log n \\
& \leq k^k \cdot {(d+1)}^{2k} \cdot e^{k - \frac{k^2}{{\ell}^2}} \cdot {\ell}^{\OO(1)} \cdot \log n \qquad \qquad (\mbox{Because } (1+p )\leq e^p, \forall p >0) \\
& = k^k \cdot {(d+1)}^{2k} \cdot e^{k - \frac{1}{{(d+1)}^2}} \cdot {(k(d+1))}^{\OO(1)} \cdot \log n  \qquad \qquad (\mbox{Because } \ell = k(d+1))
\end{align*}
This completes the proof. 
\end{proof}

\subsection{Extensions}\label{sec:ISCLemmaExtensions}

For some graphs, whose degeneracy is not bounded, it may still be possible to find a \enquote{small} sized \icf. This is captured by the Corollary \ref{cor:iscl}. 

\begin{corollary}\label{cor:iscl}
Let  $d,k\in {\mathbb N}$ and $G$ be a graph.  Let $S \subseteq V(G)$ be such that 
$G -S$ is $d$-degenerate. 
%
There are two deterministic algorithms  which 
given $d,k\in {\mathbb N}$, $G$ and $S$, run in time $\OO(2^{|S|} \cdot \binom{k(1+d)}{k} \cdot 2^{o(k(1+d))} \cdot (n+m)\log n)$ and $\OO(2^{|S|} \cdot \binom{k^2 {(1+d)}^2}{k} \cdot {(k(1+d))}^{\OO(1)} \cdot (n+m) \log n)$, 
and outputs an \icf for $(G,k)$ of size at most $2^{|S|} \cdot \binom{k(1+d)}{k} \cdot 2^{o(k(1+d))} \cdot \log n$ and $2^{|S|} \cdot \binom{k^2{(1+d)}^2}{k} \cdot {(k(1+d))}^{\OO(1)} \cdot \log n$ respectively.
\end{corollary}
\begin{proof}
Let $G'=G-S$. By the property of $S$, we know that $G'$ is $d$-degenerate. 
We first apply Lemma~\ref{lemma:discl} and get a $k$-independent set covering 
family ${\mathcal{F}}^{'}$ for $(G',k)$. Then we output the family 
$$\ifam{G}{k}=\{(A\cup B)\setminus N_G(B)~\vert~ A\in {\mathcal{F}}^{'}, B\subseteq S \mbox{ is an independent set in } G\}.$$
We claim that $\ifam{G}{k}$ is a $k$-\icf for $(G,k)$.  
Towards that, first we prove that all sets in $\ifam{G}{k}$ are independent sets in $G$. 
Let $Y\in \F$. We know that $Y=(A\cup B)\setminus N_G(B)$, for 
some  $A\in {\mathcal{F}}^{'}$ and $B\subseteq S$which is an independent set in $G$. 
By the definition of ${\mathcal{F}}^{'}$, $A$ is an independent set in $G$. 
Since $A$ and $B$ are independent sets in $G$, 
$Y=(A\cup B)\setminus N_G(B)$ is an independent set in $G$. 
Now we show that for any independent set $X$ in $G$ of cardinality 
at most $k$, there is an independent set containing $X$ in $\ifam{G}{k}$. 
Let $X=X'\uplus X''$, where $X'=X\setminus S$ and $X''=X\cap S$. By the definition 
of ${\mathcal{F}}^{'}$, there is a set $Z\in {\mathcal{F}}^{'}$ such that $X'\subseteq Z$.  
Then the set $(Z\cup X'')\setminus N_G(X'') \in \ifam{G}{k}$ is the required  independent set containing $X$. 
Observe that $\vert \ifam{G}{k} \vert \leq \vert \mathcal{F}^{'} \vert \cdot 2^{\vert S \vert}$. Also, the running time of this algorithm is equal to the time taken to compute ${\mathcal{F}}^{'}$ plus $ \vert \ifam{G}{k} \vert \cdot (n+m)$. Thus, the running time and the bound on the cardinality of $\ifam{G}{k}$ as claimed in the lemma follows from Lemmas~\ref{lemma:discl} and~\ref{lemma:discl2}.
\end{proof}

\newcommand{\nab}{\mathop{\triangledown}}
\newcommand{\td}{\mathbf{td}}
\newcommand{\idf}{\mathbf{idf}}
\newcommand{\ch}{\mathbf{child}}
\newcommand{\clos}{{\mathbf{clos}}}
\newcommand{\dist}{{\mathbf{dist}}}
\newcommand{\height}{{\mathbf{height}}}
\newcommand{\depth}{{\mathbf{depth}}}
\newcommand{\reach}{{\mathbf{reach}}}
\newcommand{\anc}{{\mathbf{anc}}}

\newcommand{\dvorakfun}{f_{\textrm{dv}}}  
\newcommand{\neifun}{f_{\textrm{nei}}}    
\newcommand{\chrgfun}{f_{\textrm{chrg}}}  
\newcommand{\gradfun}{f_{\nabla}}           
\newcommand{\wcolfun}{f_{\wcol}}            
\newcommand{\cnumfun}{f_{\numcliques}}            

\subsection{Nowhere Dense Graphs}
In this section, we give an efficient construction of $k$-independence covering families of size at most $f(k)\cdot n$ for every {\em nowhere dense} class of graphs~\cite{nevsetvril2008grad,nevsetvril2011nowhere}, based on low treedepth colorings of such graphs. 
The class of nowhere dense graphs is a common generalization of proper minor closed classes, classes of graphs with bounded degree, graph class locally excluding a fixed graph $H$ as minor and classes of bounded expansion~(see \cite[Figure~$3$]{nevsetvril2011nowhere}).

In order to define the class of nowhere dense graphs, we need several new definitions. 

\begin{definition}[Shallow minor]
  A graph~$M$ is an \emph{$r$-shallow minor} of~$G$, where~$r$ is an
  integer, if there exists a set of disjoint subsets $V_1, \ldots,
  V_{|M|}$ of~$V(G)$ such that
  \begin{enumerate}
  \setlength{\itemsep}{-2pt}
  \item each graph $G[V_i]$ is connected and has radius at most~$r$,
    and
  \item there is a bijection $\psi \colon V(M) \to \{V_1, \ldots,
    V_{|M|}\}$ such that for every edge $uv \in E(M)$ there is an edge
    in~$G$ with one endpoint in $\psi(u)$ and second in $\psi(v)$.
  \end{enumerate}
  The set of all $r$-shallow minors of a graph~$G$ is denoted by~$G
  \nab r$.  Similarly, the set of all $r$-shallow minors of all the
  members of a graph class $\cal G$ is denoted by 
  ${\cal G} \nab r =\bigcup_{G \in {\cal G}} (G \nab r)$.
\end{definition}
We first introduce the definition of a nowhere dense graph class; let $\omega(G)$ denotes the size of the largest clique in $G$ and $\omega({\cal G})=\sup_{G\in \cal G} \omega(G)$.

\begin{definition}[Nowhere dense]\label{def:nd}
  A graph class $\cal G$ is \emph{nowhere dense} if there exists a
  function $f_\omega \colon \mathbb{N} \to \mathbb{N}$ such that for
  all~$r$ we have that $\omega({\cal G} \nab r) \leq f_\omega(r)$.
\end{definition}

We will mostly rely on the low treedepth colorings of nowhere dense graph classes. Towards that, we first define 
the notion of treedepth. 

\begin{definition}[Treedepth of a graph]
A \emph{treedepth decomposition} of a graph $G$ is a rooted forest $F$ on the vertex set $V(G)$, that is $V(F)=V(G)$, such that for every edge $uv \in E(G)$, the endpoints $u$ and $v$ are in ancestor-descendant relation. The \emph{height} of a rooted forest $F$, denoted by $\height(F)$, is the maximum number of vertices on a simple path from the root of $F$ to a leaf in $F$. The \emph{treedepth} of $G$, denoted $\td(G)$, is the least $d \in \mathbb{N}$ such that there exists a treedepth decomposition $F$ of $G$ with $\height(F)=d$.
\end{definition}

\begin{proposition}\label{prop:tdtw}
Let $G$ be a graph. Then, $\dg(G) \leq \tw(G) \leq \td(G)-1$ . 
\end{proposition}

Next we define the notion of {\em treedepth colorings} and state a result that shows that nowhere dense graph classes admit  low treedepth colorings.
\begin{definition}[Treedepth coloring]
 An $r$-treedepth colouring of $G$ is a colouring such that any $r'\leq r$ color classes induce a subgraph with 
 treedepth at most $r'-1$. The minimum number of colors of such a coloring of $G$ is denoted by $\mathbf{tdcolr}_r(G)$.
\end{definition}

\begin{proposition}[Theorem 5.6,\cite{nevsetvril2011nowhere}]\label{prop:nowherepartsalgo}
Let $\cal G$ be a nowhere dense graph class. Then there is a function  $f(r,\eps)$ such that 
$\mathbf{tdcolr}_r(G)\leq f(r,\eps)\cdot n^\eps$ for every integer $r\geq 0$, $G\in \cal G$, and real $\eps>0$. Furthermore, a  $f(r,\eps)\cdot n^\eps$-treedepth colouring of $G$ can be obtained in time $\OO(f(r,\eps) \cdot n^{1+o(1)})$.
\end{proposition}

We refer readers to the book by Nesetril and Ossona de Mendez~\cite{DBLP:books/daglib/0030491} for a detailed exposure to nowhere dense classes of graphs, their alternate characterization and several properties about it. See also~\cite{GroheKS13}. 

We present two algorithms for the construction of \enquote{small} sized \icf for $(G,k)$, where $G$ belongs to the class of nowhere dense graphs. There are two core ingredients of these algorithms - the first one are the deterministic algorithms of Lemmas~\ref{lemma:discl} and ~\ref{lemma:discl2} that compute an \icf for bounded degeneracy graphs, and the second one is Proposition~\ref{prop:nowherepartsalgo} that states that for any integer $r$, the vertices of the graph belonging to the nowhere dense graph class, can be partitioned in such a way that the graph induced on any of the $i \leq r$ parts of this partition has treedepth bounded by $i$. Since bounded treedepth implies bounded degeneracy~\cite[Propositon 6.4]{DBLP:books/daglib/0030491}, we can compute \icf for each such subgraph (that has bounded treedepth) using the algorithms of Lemmas~\ref{lemma:discl} and ~\ref{lemma:discl2}. We can then combine these independent set covering families of such subgraphs to obtain an \icf for the whole graph. This whole idea is formalized in Lemma~\ref{lemma:disclnowhere}.

\begin{lemma}
\label{lemma:disclnowhere}
Let $G$ be a graph such that $G \in \mathcal{G}$, where $\mathcal{G}$ is a class of nowhere dense graphs. For any $k \in \mathbb{N}$,
there are two deterministic algorithms that run in time $$\OO\left(f(k,\frac{1}{k}) \cdot n^{1+o(1)} + g(k)\cdot \binom{k(1+d)}{k} \cdot 2^{o(k(1+d))} \cdot n(n+m)\log n\right)$$ and $$\OO\left(f(k,\frac{1}{k}) \cdot n^{1+o(1)} + g(k) \cdot \binom{k^2 {(1+d)}^2}{k} \cdot {(k(1+d))}^{\OO(1)} \cdot n(n+m) \log n\right),$$ and output a $k$-\icf for $(G,k)$ of size $\OO(g(k) \cdot \binom{k(1+d)}{k} \cdot 2^{o(k(1+d))} \cdot n \log n)$ and $\OO(g(k) \cdot \binom{k^2{(1+d)}^2}{k} \cdot {(k(1+d))}^{\OO(1)} \cdot n \log n)$ respectively, where $f$ is a function defined  in Proposition~\ref{prop:nowherepartsalgo} and $g(k) = {(f(k,\frac{1}{k}))}^k$.
\end{lemma}

\begin{proof}
For the given graph $G$, and an integer $k$, set $\epsilon=\frac{1}{k}$. Now compute the partition of $V(G)$ into
 $p = \OO(f(k,\frac{1}{k}) \cdot n^{1/k})$ parts, say $V_1, \ldots, V_p$, using the algorithm of Proposition~\ref{prop:nowherepartsalgo}. For all $A \in \binom{[p]}{k}$, let $G_A = G[\bigcup_{i \in A} V_i]$. Let ${\F}_A$ be an \icf for $(G_A,k)$. Let $\ifam{G}{k} = \bigcup_{A \in \binom{[p]}{k}} {\F}_A$. We claim that $\ifam{G}{k}$ is a $k$-\icf for $(G,k)$. Let $X$ be an independent set in $G$ of size at most $k$. We show that there exists a set $Y \in \ifam{G}{k}$ such that $X \subseteq Y$. Let $A' \subseteq [p]$ such that for any $i \in [p]$, $X \cap V_i \neq \emptyset$ if and only if $i \in A'$. Let $A \subseteq [p]$ such that $A' \subseteq A$ and $|A| = k$. Observe that $X$ is an independent set in $G_A$ and then thee exists $Y \in {\F}_A$ such that $X \subseteq Y$. Sinc $\ifam{G}{k} = \bigcup_{\binom{[p]}{k}} {\F}_A$, $Y \in \ifam{G}{k}$.

Thus, to compute $\ifam{G}{k}$, one needs to compute the partition $V_1, \ldots, V_p$ and ${\F}_A$ for each $A \in \binom{[p]}{k}$. From Proposition~\ref{prop:nowherepartsalgo}, $\td(G_A) \leq k-1$ and thus, the degeneracy of 
$(G_A) \leq k-1$ (\cite[Propositon 6.4]{DBLP:books/daglib/0030491}). Thus, from Proposition~\ref{prop:nowherepartsalgo} and, Lemmas~\ref{lemma:discl} and~\ref{lemma:discl2}, there are deterministic algorithms that run in time 
$$\OO\left(f(k,\frac{1}{k})  \cdot n^{1+o(1)} +  \binom{p}{k} \cdot \binom{k(1+d)}{k} \cdot 2^{o(k(1+d))} \cdot (n+m)\log n\right)$$ and $$\OO\left(f(k,\frac{1}{k})  \cdot n^{1+o(1)} + \binom{p}{k} \cdot \binom{k^2 {(1+d)}^2}{k} \cdot {(k(1+d))}^{\OO(1)} \cdot (n+m) \log n\right),$$ and output an \icf for $(G,k)$ of size $\OO(\binom{p}{k} \cdot \binom{k(1+d)}{k} \cdot 2^{o(k(1+d))} \cdot \log n)$ and $\binom{p}{k} \cdot \binom{k^2{(1+d)}^2}{k} \cdot {(k(1+d))}^{\OO(1)} \cdot \log n$, respectively. Here, $f$ is the function described in Proposition~\ref{prop:nowherepartsalgo}. Observe that $\binom{p}{k} \leq (f(k,\frac{1}{k}) n^{1/k})^k\leq g(k) \cdot n$, where $g(k) = {(f(k,\frac{1}{k}))}^k$. This concludes the proof. 
\end{proof}

\subsection{Barriers}\label{sec:ISCLemmaBarriers}

In this subsection we show that we can not get small independence covering families on general graphs. 
We also show that we can not get small covering families when we generalize the notion of ``independent set''  to something similar even on graphs of bounded degeneracy.  

\paragraph*{Independence covering family for general graphs.}
Let $k$ be a positive integer. Consider the graph $G$ on $n$ vertices, where $n$ is divisible by $k$, which is a disjoint collection of $k$ cliques on $\frac{n}{k}$ vertices each. Let $C_1, \ldots, C_k$ be the disjoint cliques that comprise $G$. Let $\ifam{G}{k}$ be a $k$-\icf for $(G,k)$. Then, we claim that, $\vert \ifam{G}{k} \vert \geq {\left ( \frac{n}{k} \right)}^k$. Consider the family $\mathcal{I}$ of independent sets of $G$ of size at most $k$ defined as $\mathcal{I} =\{ \{v_1, \ldots, v_k\} : \forall i \in [k], v_i \in C_i\}$. Note that $\vert \mathcal{I} \vert = {\left ( \frac{n}{k} \right)}^k$. We now prove that, it is not the case that there exists $Y \in \ifam{G}{k}$ such that for two distinct sets $X_1, X_2 \in \mathcal{I}$, $X_1, X_2  \subseteq Y$. This would imply that $\vert \ifam{G}{k} \vert \geq {\left ( \frac{n}{k} \right)}^k$. Suppose, for the sake of contradiction, that there exists $Y \in \ifam{G}{k}$ and $X_1, X_2 \in \mathcal{I}$ such that $X_1 \neq X_2, X_1 \subseteq Y$ and $X_2 \subseteq Y$. Since $X_1 \neq X_2$, there exist $u \in X_1$ and $v \in X_2$ such that $u,v \in C_i$ for some $i\in [k]$. Since $X_1 \subseteq Y$ and $X_2 \subseteq Y$, $u,v \in Y$, which contradicts the fact that $Y$
 is an independent set in $G$ (because $uv\in E(G)$).

\paragraph*{Induced matching covering family for disjoint union of stars.} 
We show that if we generalize independent set to induced matching, then we can not hope for  
small covering families even on the  disjoint union of {\em star} graphs, which are graphs of degeneracy one.   
\begin{definition}[Induced Matching Covering Family]
For a graph $G$ and a positive integer $k$, a family $\mathcal{M} \subseteq 2^{V(G)}$ is called an \emph{induced matching covering family} for $(G,k)$ if for all $Y \in \mathcal{M}$, $G[Y]$ is a matching, that is, each vertex of $Y$ has degree exactly one in $G[Y]$, and for any induced matching $M$ in $G$ on at most $k$ vertices, there exists $Y \in \mathcal{M}$ such that $V(M) \subseteq Y$.
\end{definition}

Let $k$ be a positive integer. Consider the graph $G$ on $n$ vertices, where $2n$ is divisible by $k$, which is a disjoint collection of $k$ stars 
on $\frac{2n}{k}$ vertices ($K_{1,\frac{2n}{k}-1}$).
That is each connected component of $G$ is isomorphic to $K_{1,\frac{2n}{k}-1}$. 
Let $\mathcal{R}$ be the set of all maximal matchings in $G$. Each matching in $R$ consists 
of $\frac{k}{2}$ edges, one from each connected component. Observe that all these matchings are induced matchings in $G$. Union of any two distinct matchings in $\mathcal{R}$ will have a $P_3$. This implies that 
 the cardinality of any induced matching covering family for $(G,k)$ is at least $\vert \mathcal{R}\vert = (\frac{2n}{k}-1)^{\frac{k}{2}}$.

\paragraph{$r$-scattered covering family for disjoint union of stars.}
Let $G$ be a graph. For any $r \in \mathbb{N}$, $X \subseteq V(G)$ is called an $r$-{\em scattered set} in $G$, if for any $u,v \in V(G)$, $d_G(u,v) > r$. An independent set in $G$ is a $1$-scattered set in $G$. 

\begin{definition}[$r$-scattered Covering Family]
For any $r \in \mathbb{N}$, for a graph $G$ and a positive integer $k$, a family $\mathcal{S} \subseteq 2^{V(G)}$ is called a \emph{$r$-scattered covering family} for $(G,k)$ if for all $Y \in \mathcal{S}$, $Y$ is an $r$-scattered set in $G$ and for any $X \subseteq V(G)$ of size at most $k$ such that $X$ is an $r$-scattered set in $G$, there exists $Y \in \mathcal{S}$ such that $X \subseteq Y$.
\end{definition}

Let $k$ be a positive integer. Consider the graph $G$ on $n$ vertices, where $n$ is divisible by $k$, which is a disjoint collection of $k$ stars 
on $\frac{n}{k}$ vertices ($K_{1,\frac{n}{k}-1}$). That is each connected component of $G$ is isomorphic to $K_{1,\frac{n}{k}-1}$. 
Notice that $G$ is a $1$-degenerate graph. 
Let $C_1,\ldots , C_k$ be the components of $G$. Define
$\mathcal{I} =\{ \{v_1, \ldots, v_k\} : \forall i \in [k], v_i \in C_i\}$. Clearly each set in $\I$ is a $r$-scattered set for any $r\in {\mathbb N}$. 
Moreover, union of any two distinct sets in $\I$ is not a $2$-scattered set. 
This implies that  the cardinality of any $r$-scattered covering family for $(G,k)$ is at least $\vert {\mathcal I} \vert =\left( \frac{n}{k}\right)^k$ for 
any $r\geq 2$.   

\paragraph{Acyclic covering family for 2-dgenerate graphs.}
We show that covering families for induced acyclic subgraphs on $2$-degenerate graphs will have large cardinality.  
\begin{definition}[Acyclic Set Covering Family]
For a graph $G$ and a positive integer $k$, a family $\mathcal{A} \subseteq 2^{V(G)}$ is called an \emph{acyclic set covering family} for $(G,k)$ if for all $Y \in \mathcal{M}$, $G[Y]$ is a forest and for any $X \subseteq V(G)$ of size at most $k$ such that $G[X]$ is a forest, there exists $Y \in \mathcal{A}$ such that $X \subseteq Y$.
\end{definition}

Let $k$ be a positive integer. Consider the graph $G$ on $n$ vertices, where $3n$ is divisible by $k$, 
which is a disjoint union of $\frac{k}{3}$ complete bipartite graphs $K_{2, \frac{3n}{k}}$. 
The degeneracy of $G$ is $2$. 
Without loss of generality assume that $\frac{3n}{k}$ is strictly more than $2$. Let $H_1,\ldots,H_{\frac{k}{3}}$ 
be the connected components of $G$. Let $H_i=(L_i\uplus R_i,E_i)$, where $\vert L_i \vert=2$. 
Now consider the family of sets ${\mathcal I}=\{L_1\cup\ldots \cup L_{\frac{k}{3}}\cup \{v_1,\ldots, v_{\frac{k}{3}}\}~|~v_i\in R_i\}$. 
Each set in $\I$ induces a collection of induced paths on $3$ vertices ($P_3$). Also, union of any two sets in $\I$ contains a cycle 
on $4$ vertices and hence, not acyclic. This implies that the cardinality of any acyclic set covering family for $(G,k)$ is at least 
$\vert {\mathcal I}\vert=\left( \frac{3n}{k}-2\right)^{\frac{k}{3}}$.

\section{Applications I: Degenerate Graphs}\label{sec:AppI}
In this section we give \FPT algorithms  for {\sc Stable} $s$-$t$ {\sc Separator}, {\sc Stable Odd Cycle Transversal}, 
{\sc Stable Multicut}  and for {\sc Stable Directed Feedback Vertex Set} on $d$-degenerate graphs, by applying 
Lemmas~\ref{lem:indcover}, \ref{lemma:discl} and  \ref{lemma:discl2}. All these algorithms, except the one for  {\sc Stable Directed Feedback Vertex Set}, are later used as a subroutine to design \FPT algorithms on general graphs.

\newcommand{\struntime}{\binom{k(1+d)}{k} \cdot k^2(1+d)}
\newcommand{\sttimes}{\binom{k(1+d)}{k} \cdot k(1+d)}
\newcommand{\stdet}{\OO(\binom{k^2{(1+d)}^2}{k} \cdot {(k(1+d))}^{\OO(1)} \cdot (n+m)\log n)}
\newcommand{\stdetfirst}{\OO(\binom{k(1+d)}{k} \cdot 2^{o(k(1+d))} \cdot k \cdot (n+m) \log n)}

\subsection{Stable $s$-$t$-Separator}\label{sec:AppIst} 
In this subsection, we study the problem of \istsfull{} 
on graphs of bounded degeneracy. The problem is formally defined below.

\medskip
\defparproblem{\istsfull{} (\ists)}{A graph $G$, $s,t \in V(G)$ and $k\in {\mathbb N}$.}{$k$}
{Is there an \sep{s}{t} $S$ in $G$ of size at most $k$ such that $S$ is an independent set in $G$?}
\medskip

In \cite{marx2013finding}, the authors showed that \ists{} is \FPT by giving an algorithm that runs in time, that is roughly, $2^{2^{k^{\OO(1)}}} \cdot n \cdot \alpha(n,n)$, where $\alpha$ is the inverse Ackermann function. In this section, we give improved algorithm for \ists{} when the input graph is $d$-degenerate, for some non-negative integer $d$. In particular, we give a randomized algorithm for \ists{}, when the input graph is $d$-degenerate, that runs in time 
$\OO(\struntime \cdot (n+m))$
and a deterministic algorithm for the same that runs in time 
$\min \{\stdetfirst,\stdet\}$.  

The core idea behind both the algorithms is that, with an independent set covering family for $(G,k)$ at hand, the independent solution to the problem lies inside one of the set in this family. Thus, instead of looking for an independent \sep{s}{t} separator in a graph, one can look for an \sep{s}{t} separator that is contained inside one of the sets in this family.
To shave off the log factor in the randomized algorithm, that we would get if we construct an independent set covering family using the algorithm of Lemma~\ref{lemma:riscl}, we use \Cref{alg:onerandomprocess} 
in our algorithm instead of constructing the whole \ifam{G}{k} before hand using multiple rounds of \Cref{alg:onerandomprocess}. 
We now define the annotated verison of the \sep{s}{t} separator problem, which will eventually be the core problem that we would be required to solve, in order to give an algorithm for \ists.


\medskip
\defparproblem{\astsfull{} (\asts)}{A graph $G$, $s,t \in V(G)$, $Y \subseteq V(G)$ and $k\in {\mathbb N}$.}{$k$}{Is there an \sep{s}{t} $S$ of size at most $k$ in $G$ such that $S\subseteq Y$?}



\begin{lemma}\label{lemma:asts}
\asts{} can be solved in time $\OO(k \cdot (n+m))$. 
\end{lemma}
\begin{proof}[Proof Sketch]
To prove the lemma, we apply \Cref{prop:ststime} on 
$(G,s,t,w, k)$, where $w$ is defined as follows: $w(v)=1$ if $v\in Y$ and $k+1$ 
otherwise. 
\end{proof}

\begin{theorem}\label{thm:istsran} 
There is a randomized algorithm which solves \ists{} on $d$-degenerate graphs 
with a worst case running time of 
$\OO(\struntime \cdot (n+m))$. 
If the input is a \yes\ instance, then the algorithm output \yes\ with  
probability at least $1-1/e$ and if it is a \no\ instance, then the algorithm always outputs \no. 
\end{theorem}
\begin{proof}
Our algorithm runs the following two step procedure 
$\sttimes$ 
many times. 
\begin{enumerate}
\item Run \Cref{alg:onerandomprocess} on $(G,k)$ and let $Z$ be its output. 
\item \label{step2st} Run the algorithm of \Cref{lemma:asts} on the instance $(G,s,t,k,Z)$ of \asts. 
\end{enumerate}
Our algorithm will output \yes, if Step 2 
returns \yes\ at least once. Otherwise, our algorithm will output \no. 
We now prove the correctness of our algorithm. Since, in Step 1, the output set $Z$ is always an independent set of $G$, 
if the algorithm returns \yes, the input instance is a \yes\ instance. For the other direction, suppose the input instance 
is a \yes\ instance. Let $X$ be a solution to it. Since $X$ is an independent set, from \Cref{lem:indcover}, $X\subseteq Z$ with probability at least $p=\succprob$. Thus, the probability that in all the executions of Step 
$1$, $X \not \subseteq Z$ is at most $(1-p)^{1/p}\leq 1/e$. Therefore, the probability that in at least one execution 
of Step 1, $X\subseteq Z$, is at least $1-1/e$. Now, consider the iteration of the algorithm when $X \subseteq Z$. For this iteration, $(G,s,t,k,Z)$ is a \yes\ instance of \asts, and thus, our algorithm will output \yes\ in this iteration. Therefore, if the input instance is a \yes/ instance, our algorithm will output \yes\ with 
probability at least $1-1/e$.  

The running time of our algorithm follows from \Cref{lem:indcover,lemma:asts}. 
\end{proof}

\begin{theorem}\label{thm:ists}
There is a deterministic algorithm which solves \ists{} on $d$-degenerate graphs in  
 time 
$\min \{\stdetfirst,\stdet\}$. 
\end{theorem}
\begin{proof}[Proof Sketch]
Let \ifam{G}{k} be an independent set covering family obtained by the algorithm of \Cref{lemma:discl} or \Cref{lemma:discl2}. 
Our algorithm runs Step $2$ of the procedure described in the proof of \Cref{thm:istsran} for all $Z\in \ifam{G}{k}$. It outputs \yes\ if at least one execution 
gives a \yes\ answer. Otherwise, it outputs \no. The correctness of the algorithm follows 
from the definition of independent set covering family and \Cref{lemma:asts}. 
The running time of the algorithm follows from \Cref{lemma:discl,lemma:discl2,lemma:asts}. 
\end{proof}

\newcommand{\octrruntime}{4^k \cdot \binom{k(1+d)}{k} \cdot k^7(1+d)}
\newcommand{\octtimes}{\binom{k(1+d)}{k} \cdot k(1+d)}
\newcommand{\coctdet}{ \OO(\binom{  4^k \cdot k^2{(1+d)}^2}{k} \cdot {(k(1+d))}^{\OO(1)} \cdot (n+m)\log n)}
\newcommand{\coctdetfirst}{\OO(4^k \cdot  k^6   \cdot \binom{k(1+d)}{k}  \cdot 2^{o(k(1+d))} \cdot (n+m) \log n) }

\subsection{Stable Odd Cycle Transversal}\label{sec:AppIOCT}

In this section, we study the problem of \ioctfull{} (also named as \stablebip{} in \cite{marx2013finding}) where the input graph has bounded degeneracy. The problem is formally defined below.

\medskip
\defparproblem{\ioctfull{} (\ioct)}{A graph $G$ and $k\in {\mathbb N}$.}{$k$}{Is there a set $X \subseteq V(G)$  of size at most $k$ such that $X$ is an independent set in $G$ and $G \minus X$ is acyclic?}
\medskip


In \cite{marx2013finding}, the authors showed that \ioct{} is \FPT by giving an algorithm that runs in time 
$2^{2^{k^{\OO(1)}}} \cdot n \cdot \alpha(n,n)$, where $\alpha$ is the inverse Ackermann function. In this section, we give an improved algorithm for \ioct{} when the input graph is $d$-degenerate, for some non-negative integer $d$. In particular, we give a randomized algorithm for \ioct{} on 
 $d$-degenerate graphs that runs in time 
$\OO(\octrruntime \cdot (n+m))$ 
and a deterministic algorithm for the same that runs in time 
$\min \{ \coctdetfirst, \coctdet \}$. As was the case with \ists, 
the core idea behind both the algorithms is that, with an independent set covering family for $(G,k)$ at hand, the independent solution to the problem lies inside one of the set in this family. 
We now define the annotated verison of the \oct{} problem.

\medskip
\defparproblem{\aoctfull{} (\aoct)}{A graph $G$, $Y \subseteq V(G)$, and $k\in {\mathbb N}$.}{$k$}{Is there a set $X \subseteq Y$ of size at most $k$ such that $G \minus X$ is acyclic?}
\medskip

When $Y=V(G)$ in \aoct, the problem is well-known by the name of \octfull{}(\oct). We will need the following result about \oct. 
\begin{proposition}[\cite{ramanujan2014linear}]\label{prop:solveoct}
\oct{} can be solved in time $\OO(4^k \cdot k^4 \cdot (n+m))$. 
\end{proposition}

Using \Cref{prop:solveoct}, we can get the following result about \aoct. 
\begin{lemma}
\label{lem:acotred}
\aoct\ can be solved in time $\OO(4^k \cdot k^6 \cdot (n+m))$
\end{lemma}

\begin{proof}[Proof sketch]
We give a polynomial time reduction from \aoct\ to \oct\ as follows. 
We replace each $v\in V(G)\setminus Y$, with $k+1$ vertices $v_1,\ldots v_{k+1}$ 
with same neighbourhood as $v$, that is, the neighbourhood of $v_1,\ldots v_{k+1}$ are same in the resulting 
graph (see \Cref{fig:red} for an illustration). Let $G'$ be the resulting graph. Then any minimal odd cycle transversal  which contain a vertex from 
$\{v_1,\ldots,v_{k+1}\}$ will also 
contain all the vertices in $\{v_1,\ldots,v_{k+1}\}$. 
Thus to find a $k$ sized solution for \aoct , 
it is enough to find an odd cycle transversal of size $k$ in $G'$. 
The total number of vertices in $G'$ is at most $k\vert V(G)\vert$ and the 
total number of edges in $G'$ is at most $(k+1)^2 \vert E(G)\vert$. 
Thus the running time of the algorithm follows from \Cref{prop:solveoct}. 
\end{proof}


\begin{figure}[t]
\centering
\begin{subfigure}[b]{0.35\textwidth}
\begin{tikzpicture}[scale=1.2]
\node [] (a) at (0,0) {$\bullet$};
\node [] (a) at (0,2) {$\bullet$};
\node [] (a) at (3,2) {$\circ$};
\node [] (a) at (3,0) {$\circ$};
\draw (0,0)--(0,2)--(2.93,2);
\draw (3,1.97)--(3,0.03);
\draw(0,0)--(2.93,0);
\end{tikzpicture}
        \end{subfigure}%
        \begin{subfigure}[b]{0.35\textwidth}
\begin{tikzpicture}[scale=1]
\node [] (a) at (0,0) {$\bullet$};
\node [] (a) at (0,2) {$\bullet$};
\draw (3,2.5) circle (0.8cm);
\draw (3,0.5) circle (0.8cm);
\node [] (a) at (3,2) {$\circ$};
\node [] (a) at (2.6,2.5) {$\circ$};
\node [] (a) at (3.1,3) {$\circ$};
\node [] (a) at (3.5,2.4) {$\circ$};
%
\node [] (a) at (3,0) {$\circ$};
\node [] (a) at (2.6,0.5) {$\circ$};
\node [] (a) at (3.1,1) {$\circ$};
\node [] (a) at (3.5,0.4) {$\circ$};
\draw 
(0,0)--(0,2);
\draw[line width=0.5mm,blue] (0,0)--(2.2,0.6);
\draw[line width=0.5mm,blue] (0,2)--(2.2,2.6);
\draw[line width=0.5mm,blue] (3,1.72)--(3,1.3);
\end{tikzpicture}
        \end{subfigure}%
        \caption{The graph at the right hand side is obtained by the reduction on the graph at the left hand side, where $k=3$ and $Y$ is the set of black colored vertices. Thick lines represents all possible edges between two sets of vertices.}\label{fig:red}
\end{figure}
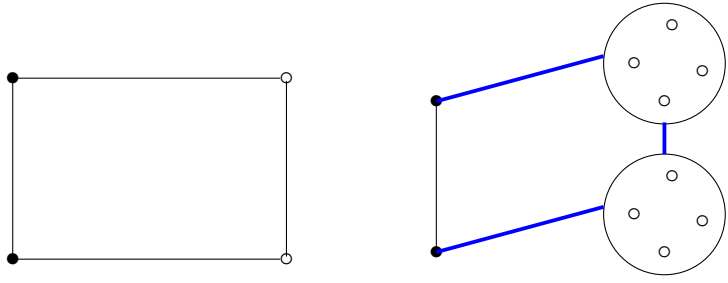

By applying \Cref{lem:acotred}, instead of \Cref{lemma:asts}, in \Cref{thm:istsran,thm:ists}, we get 
the following theorems.  

\begin{theorem}\label{thm:ioctran} 
There is a randomized algorithm which solves \ioct\ on $d$-degenerate graphs 
with a worst case running time of
$\OO(\octrruntime \cdot (n+m))$. 
If the input is a \yes\ instance, then the algorithm output \yes\ with  
probability at least $1-1/e$ and if it is a \no\ instance, then the algorithm always outputs \no. 
\end{theorem}

\begin{theorem}\label{thm:ioctdet} 
There is a deterministic algorithm which solves \ioct\ on $d$-degenerate graphs 
in time $\min \{\coctdetfirst, \coctdet\}$. 
\end{theorem}

\newcommand{\idfvstimes}{\OO((k+1)! \cdot 4^k \cdot \binom{k(1+d)}{k} \cdot k^6(1+d) \cdot (n+m))}
\newcommand{\idfvstimedet}{\OO((k+1)! \cdot 4^k \cdot \binom{k^2{(1+d)}^2}{k} \cdot {(k(1+d))}^{\OO(1)} \cdot (n+m)\log n)}
\newcommand{\idfvstimedetfirst}{\OO( (k+1)! \cdot 4^k \cdot k^5 \cdot \binom{k(1+d)}{k} \cdot 2^{o(k(1+d))} \cdot (n+m) \log n  )}
\subsection{Stable Directed Feedback Vertex Set}
\label{sec:AppIDFVS}

In this section, we study the problem of \idfvsfull{} (\idfvs). \idfvs{} was shown to be {\sc W[1]-Hard} in \cite{misra2012parameterized}.  We study \idfvs{} restricted to the case where the input graph has bounded degeneracy. The problem is formally defined below.

\medskip
\defparproblem{\idfvsfull{} (\idfvs)}{A digraph $D$ and  $k\in {\mathbb N}$.}{$k$}{ Is there a set $X \subseteq V(D)$ of size at most $k$ such that $S$ is an independent set in $D$ and $D\minus S$ is a directed acyclic graph?} 
\medskip

As with the algorithms in the previous sections, the algorithm for this problem follows the same outline.

We need to use the known algorithm for \dfvsfull{} (\dfvs) - the same problem as \idfvs{} where the solution need not be an independent set. 
\begin{lemma}[\cite{LokshtanovRS16}]\label{prop:dfvs}
\dfvs{} can be solved in time $\OO((k+1)! \cdot 4^k \cdot k^5 \cdot (n+m))$. 
\end{lemma}

\begin{theorem}\label{thm:idfvsran} 
There is a randomized algorithm which solves \idfvs{} on $d$-degenerate graphs 
with a worst case running time 
$\idfvstimes$. 
If the input is a \yes\ instance, then the algorithm outputs \yes\ with  
probability at least $1-1/e$ and if it is a \no\ instance, then the algorithm always outputs \no. 
\end{theorem}
\begin{proof}
The algorithm runs the following two step procedure 
$\sttimes$ 
many times. 
\begin{enumerate}
\item Run \Cref{alg:onerandomprocess} on $(G,k)$ and let $Z$ be its output. 
\item Construct $G'$ as in \Cref{lem:acotred}, that is, add $k+1$ copies for each vertex in $V(G)\setminus Z$ to the graph $G$ such that all of them have the same neighborhood in the resulting graph. Then apply \Cref{prop:dfvs} on $(G',k)$. 
\end{enumerate}
The proof of correctness of this algorithm is similar in arguments to the proofs of \Cref{lem:acotred} and \Cref{thm:istsran}.  

The running time of the algorithm follows from \Cref{lem:indcover,prop:dfvs}. 
\end{proof}

By arguments similar to the proof  of  \Cref{thm:ists}, one can prove the following theorem.

\begin{theorem}\label{thm:idfvs}
There is a deterministic algorithm which solves \idfvs{} on $d$-degenerate graphs in  
 time 
$\min \{\idfvstimedetfirst, \idfvstimedet\}$. 
\end{theorem}

\newcommand{\imctime}{2^{\OO(k^3)} \cdot \binom{k^2{(1+d)}^2}{k} \cdot {d}^{\OO(1)} \cdot  mn \log^2 n}
\newcommand{\imctimefirst}{ 2^{\OO(k^3)} \cdot \binom{k(1+d)}{k}  \cdot 2^{o(k(1+d))} \cdot mn \log^2 n }
\newcommand{\amcfull}{\textsc{Annotated Multicut}}
\newcommand{\amc}{\textsc{AMC}}
\subsection{Stable Multicut}\label{sec:AppImulticut}

For a graph $G$ and a set of terminal pairs $T= \{\{s_1,t_1\}, \ldots, \{s_p,t_p\}\}$, $S \subseteq V(G)$ is a \emph{multicut} of $T$ in $G$ if $G \minus S$ has no path from $s_i$ to $t_i$ for any $i \in [p]$.
We say that 
$S$ is an independent multicut of $T$ in $G$, 
if $S$ is an independent set in $G$ and $S$ is a multicut of $T$ in $G$. 
In this section we prove that the problem of finding an independent multicut (formally defined below) in a bounded degeneracy graph is \FPT\ when parameterized by the solution size. 

\medskip
\defparproblem{\imc}{An undirected graph $G$, a set  of terminal pairs $T$ and $k\in {\mathbb N}$.  }
{$k$}{Is there an independent multicut of $T$ in $G$ of size at most $k$?}
\medskip

Using \Cref{lemma:discl,lemma:discl2}, and  the known  algorithm for \mc~\cite{Marx06i,LokshtanovRS16} (the problem where 
we do not demand that the multicut to be an independent set), we prove 
the main theorem of this section. Before that, we state an algorithmic result for \mc{} that is crucuially used by our algorithm.  

\begin{lemma}[\cite{Marx06i,LokshtanovRS16}]
\label{lem:marxalgo}
\mc\ can be solved in $2^{\OO(k^3)} \cdot mn\log n$ time.
\end{lemma}

We now define the annotated version of the \mc\ problem, the way we defined it for the previusly considered problems.

\medskip
\defparproblem{\amcfull}{An undirected graph $G$, a set of terminal pairs $T$, $Y \subseteq V(G)$ and $k\in {\mathbb N}$.}
{$k$}{Is there a multicut $S$ of $T$ in $G$ of size at most $k$ such that $S \subseteq Y$?}
\medskip

The following lemma give an algorithm for solving \amcfull\ using the algorithm of Lemma~\ref{lem:marxalgo}.
\begin{lemma} 
\label{lem:subsetmcalgo1}
\amcfull{} can be solved in time $2^{\OO(k^3)} \cdot mn \log n$.

\end{lemma}
\begin{proof}[Proof sketch]
We first give a polynomial time reduction from \amcfull{} to \mc which is described below. 

Let $(G,T,Y,k)$ be an instance of \amcfull. Construct a graph $G'$ from $G$ by replace each $v\in V(G)\setminus Y$, with $k+1$ vertices $v_1,\ldots v_{k+1}$ 
with same neighbourhood as $v$. That is, the neighbourhood of $v_1,\ldots v_{k+1}$ are same in the resulting 
graph $G'$. 
We call the set of vertices that are added for $v$ in $G'$ as the block for $v$. 
We now construct the set of terminal pairs $T'$ from the set of terminals $T$ as follows. 
If $\{s,t\}\in T$ and $\{s,t\}\subseteq Y$, we add $\{s,t\}$ to $T'$. 
Suppose $\{s,t\}\in T$ and $\{s,t\}\cap Y=\{t\}$. 
Let $s_1,\ldots, s_{k+1}$ be the bock for $s$ in $G'$. We add $\{s_1,t\},\ldots \{s_{k+1},t\}$ to $T'$.  
Suppose $\{s,t\}\in T$ and $\{s,t\}\subseteq V(G)\setminus Y$. 
Let $s_1,\ldots, s_{k+1}$ 
and $t_1,\ldots,t_{k+1}$
be the blocks for $s$ and $t$, respectively. We add $\{\{s_i,t_j\}~|~i,j\in [k+1]\}$ to $T'$.  

We will now show that $(G,T,Y,k)$ is a \yes{} instance of \amcfull{} if and only if $(G',T',k)$ is a \yes{} instance of \mc.
For the forward direction, let $C$ be a multicut of size at most $k$ in $G$ such that $C\subseteq Y$. 
We claim that $C$ is a multicut of $T'$ in $G'$. Suppose not. Then, there is a path from $s'$ to $t'$ in $G' \minus C$, where $\{s',t'\}\in T'$.  
Let $s$ and $t$ be the vertices in $V(G)$ such that $s'$ and $t'$ are the vertices corresponding 
to them, respectively, that is, if $s'\in Y$, then $s=s'$, otherwise let $s$ be the vertex such that $s'$ is in the block 
of vertices constructed for the replacement of $s$ in $G'$. 
By replacing 
each vertex in the $s'-t'$ path in $G'$ by the corresponding vertex in $G$, we get a walk from $s$ to $t$ in 
$G\minus C$, which contradicts the fact that $C$ is a multicut of $T$ in $G$. 
For the backward direction, suppose $C'$ is a minimal multicut of $T'$ in $G'$ of size at most $k$. Since, for any $v\in V(G)\setminus Y$, 
the neighbourhood of $v_1,\ldots v_{k+1}$ in $G'$ is the same as that of $v$ in $G$ and $\vert C ' \vert \leq k$, $C'\cap \{v_1,\ldots,v_{k+1}\}=\emptyset$. Thus, $C' \subseteq Y$. Since $G'$ is a supergraph of $G$ and $T \subseteq T'$, $C'$ is a multicut of $T$ in $G$.


Thus, to find a $k$ sized multicut of $T$ in $G$ which is fully contained in $Y$, it 
is enough to find a multicut of $T'$ in $G'$. 
The total number of vertices in $G'$ is at most $k\vert V(G)\vert$ and the 
total number of edges in $G'$ is at most $(k+1)^2 \vert E(G)\vert$. 
Thus, the running time of the algorithm follows from Lemma~\ref{lem:marxalgo}. 
This completes the proof sketch of the lemma. 
 \end{proof}

\begin{theorem}
\label{thm:imc}
\imc\ can be solved in time 
$$\min \{\imctimefirst, \imctime\}.$$ 
\end{theorem}
\begin{proof}
Let $(G,T,k)$ be an instance of \imc. 
The algorithm first computes an independent set covering family \ifam{G}{k} using the algorithm of \Cref{lemma:discl} or \Cref{lemma:discl2}. By the definition 
of independent set covering family, any independent set $S$ of $G$ of cardinality at most 
$k$ is contained in an independent set $I\in \ifam{G}{k}$. In particular, if $(G,T,k)$ is a \yes\ instance, then there 
is a solution which is fully contained in some $I \in \ifam{G}{k}$ (moreover, the set $I$ is independent in $G$). 
Therefore, to test whether $(G,T,k)$ is a \yes\ instance or not, it is enough to test whether there is multicut (not necessarily independent) of size 
at most $k$ contained in some $I \in \ifam{G}{k}$. Hence, for each $I\in \ifam{G}{k}$, the algorithm runs the algorithm of \Cref{lem:subsetmcalgo1}, 
and check whether there is a multicut of size at most $k$ in $I$ or not. If even one application of the algorithm of \Cref{lem:subsetmcalgo1} 
gives a positive answer, the algorithm outputs \yes, otherwise it outputs \no. 
The running time of this algorithm follows from \Cref{lemma:discl,lemma:discl2,lem:subsetmcalgo1}. 
\end{proof}

\section{Applications II: General Graphs}\label{sec:AppII}
In this section, we solve \istsfull{} and \ioctfull{} on general graphs. The core of our algorithms is the Treewidth Reduction Theorem of \cite{marx2013finding} and our algorithms for \ists{} and \ioct{} on bounded degeneracy graphs from \Cref{sec:AppIst,sec:AppIOCT},  respectively.
We begin by stating the Treewidth Reduction Theorem.

\begin{theorem}[Treewidth Reduction Theorem, Theorem 2.15 \cite{marx2013finding}]\label{prop:twred}
Let $G$ be a graph, $T \subseteq V(G)$ and $k\in {\mathbb N}$. Let $C$ be the set of all vertices of $G$ participating in a minimal  \sep{s}{t} of cardinality at most $k$ for some $s,t \in T$. For every $k$ and $|T|$, there is an algorithm that computes a graph $G^{*}$ having the following properties, in time $2^{{(k+|T|)}^{\OO(1)}} \cdot (n+m)$.
\begin{enumerate}
\setlength{\itemsep}{-2pt}
 \item $C \cup T \subseteq V(G^{*})$,
 \item for every $s,t \in T$, a set $K \subseteq V(G^{*})$ with $|K| \leq k$ is a minimal \sep{s}{t} of $G^{*}$ if and only if $K \subseteq C \cup T$ and $K$ is a minimal \sep{s}{t} of $G$,
 \item the treewidth of $G^{*}$ is at most $2^{({k+|T|)}^{\OO(1)}}$, and 
 \item $G^{*}[C \cup T]$ is isomorphic to $G[C \cup T]$.
\end{enumerate}
\end{theorem}
We remark here that Theorem 2.1 in \cite{marx2013finding} does not state explicit dependency on $k$ and $|T|$ in both, the running time of the algorithm 
and the treewidth of $G^{*}$ obtained. 



\paragraph*{Stable $s-t$ Separator}
Let $(G,k)$ be an instance of \ists. 
To solve \ists{} on general graphs, we first apply the Treewidth Reduction Theorem (\Cref{prop:twred}) 
on $G,T=\{s,t\}$ and $k$ to obtain a graph $G^{*}$ with treewidth is upper bounded by . We then show that for \ists{}, it is enough to work with this new graph $G^{*}$ whose treewidth is bounded $2^{{k}^{\OO(1)}}$. 
By conditions $2$ and $4$, to find a minimal independent \sep{s}{t} separator in $G$, it is enough to 
a minimal independent \sep{s}{t}  in $G^{*}$. By \Cref{prop:tdtw}, we know that the degeneracy 
of  $G^{*}$ is at most $2^{{k}^{\OO(1)}}$, and hence we apply \Cref{thm:istsran} or \Cref{thm:ists} to get a solution of  \ists{} on $(G,k)$. 
That is, we get the following theorem. 

\begin{theorem}\label{thm:genst}
There is a randomized algorithm that solves \ists{} in time $2^{k^{\OO(1)}} (n+m)$ with success probability at least $1 - \frac{1}{e}$. There is a deterministic algorithm that solves \ists{} in time $2^{k^{\OO(1)}}(n+m) \log n$. 
\end{theorem}

%


\paragraph*{Stable Odd Cycle Transversal} 
By using the \Cref{thm:genst,prop:solveoct} we get a $2^{k^{\OO(1)}}(n+m)$ time (\FPT\ linear time) algorithm 
for \ioct.  Towards that, in the Theorem $4.2$ of Marx et al.~\cite{MarxR14} we replace the algorithm of Kawarabayashi and Reed~\cite{KawarabayashiR10} with \Cref{prop:solveoct} and the algorithm for  \ists\ with  \Cref{thm:genst}. 
For completeness we include the proof here.

\begin{proposition}[Lemma 4.1, \cite{marx2013finding}]\label{prop:octsep}
Let $G$ be a bipartite graph and let $(B',W')$ be a proper $2$-coloring of the vertices. Let $B$ and $W$ be two subsets of $V(G)$. Then, for any $S \subseteq V(G)$, the graph $G \minus S$ has a $2$-coloring where $B \setminus S$ is black and $W \setminus S$ is white if and only if $S$ separates $X := (B \cap B') \cup (W \cap W')$ and $Y:=(B \cap W') \cup (W \cap B')$. 
\end{proposition}


\begin{theorem}\label{thm:genoct}
There is a randomized algorithm that solves \ioct{} in time $2^{k^{\OO(1)}} (n+m)$ with success probability at least $1 - \frac{1}{e}$. There is a deterministic algorithm that solves \oct{} in time $2^{k^{\OO(1)}}(n+m) \log n$. 
\end{theorem}

%
%

\begin{proof}
Using the algorithm of \Cref{prop:solveoct}, find a set $S_0 \subseteq V(G)$ of size at most $k$ such that $G \setminus S_0$ is a bipartite graph. Observe that if such a set does not exist then $(G,k)$ is \no{} instance of \ioct. Thus, henceforth, we can assume that such a set $S_0$ exists. We next branch into $3^{|S_0|}$ cases, where each branch has the following interpretation. If we fix a hypothetical solution $S$ and a proper $2$-coloring of $G \minus S$, then each vertex of $S_0$ is either removed (that is, belongs to $S$, a fixed hypothetical solution), colored with the first color, say black, or colored with the secong color, say white. For a particular branch, let $R$ be the the vertices of $S_0$ to be removed (in order to get the hypothetical solution $S$) and let $B_0$ (respectively $W_0$) be the vertices of $S_0$ getting color black (respectively white) in a proper $2$-coloring of $G \minus S$. A set $S$ is said to be compatible with the partition $(R,B_0,W_0)$, if $S \cap S_0 = R$ and $G \setminus S$ has a proper $2$-coloring, with colors black and white, where the vertices in $B_0$ are colored black and the vertices in $W_0$ are colored white. 
Observe that $(G,k)$ is a \yes{} intance of \ioct, if and only if for at least one branch corresponding to a partition $(R,B_0,W_0)$ of $S_0$, there is a set $S$ compatible with $(R,B_0,W_0)$ of size at most $k$ and $S$ is an independent set.  
Note that we need to check only those branches corresponding to the partition $(R,B_0,W_0)$ where $G[B_0]$ and $G[W_0]$ are edgeless graphs. 

The next step is to transform the problem of finding a set compatible with $(R, B_0,W_0)$ 
into a separation problem. Let $(B',W')$ be a $2$-coloring of $G\minus S_0$. Let $B= N(W_0) \setminus S_0$ and $W = N(B_0) \setminus S_0$. Let $X$ and $Y$ be the sets as defined in \Cref{prop:octsep}. That is, $X= (B \cap B') \cup (W \cap W')$ and $Y= (B \cap W') \cup (W \cap B')$. Construct a graph $G'$ that is obtained from $G$  by deleting the set $B_0 \cup W_0$, adding a new vertex $s$ adjacent with $X \cup R$ and adding a new vertex $t$ adjacent with $Y \cup R$. 
Notice that every \sep{s}{t} in $G'$ contains R. By \Cref{prop:octsep}, a set $S$ is compatible with $(R,B_0,W_0)$ if and only if $S$ is an 
$s-t$ separator in $G$. Thus, we need to decide whether there is an \sep{s}{t} $S$ of size at most $k$ such that $G'[S] = G[S]$ is an edgeless graph and this step can be done by \Cref{thm:genst}.

Towards the run time analysis, we run the algorithm of \Cref{prop:solveoct} once, which takes time $2^{\OO(k)}(m+n)$. Then we apply \Cref{thm:genst} at most $3^k$ times. Thus, we get the required running time. 
\end{proof}

\section{Tool II:  Multicut Covering Graph Sparsification}
\label{sec:thmone}
\newcommand{\rbound}{64^{k+1}(k+1)^2}
\newcommand{\cbound}{16^{k} \cdot 64(k+1)}
\newcommand{\dpcfull}{\textsc{Digraph Pair Cut}}



This section starts by showing how to efficiently find some vertices that are irrelevant to ``small'' digraph pair cuts (defined in \Cref{subsec:dpc}), assuming that the input graph has a sufficiently large number of vertices that are in-neighbors of the root.
Afterwards,
having a method to identify such irrelevant vertices at hand, we develop (in \Cref{subsect:multtool}) an efficient algorithm that given a graph 
$G$, a set of terminal pairs $T$ and a positive integer $k$, outputs 
an induced subgraph $G^\star$ of $G$ and a subset $T^\star\subseteq T$
such that the following conditions are satisfied. First, any set $S\subseteq V(G)$ of size at most $k$ is a minimal multicut of $T$ in $G$ if and only if $S\subseteq V(G^\star)$ and it is a minimal multicut of $T^\star$ in $G^\star$. 
Second, $G^\star$ does not contain any ``large'' $(k+2)$-connected set. Using this algorithm, we later give an \FPT\ algorithm for \imc\ on general graphs. 

\subsection{Vertices Irrelevant to Digraph Pair Cuts}
\label{subsec:dpc}
The notion of a digraph pair cut was defined by Kratsch and Wahlstr{\"{o}}m in~\cite{KratschW12}. This notion was used to derive randomized polynomial kernels for 
many problems, including {\sc Almost $2$-SAT} and {\sc Multiway Cut with Deletable Terminals}. 
Towards defining which vertices are irrelevant to ``small'' digraph pair cuts, we first formally define what is a digraph pair cut. 
\begin{definition}
Let $D$ be a digraph, $T$ be a set of pairs of vertices (called {\em terminal pairs}), 
and $r\in V(D)$. We say that $S\subseteq V(D)\setminus \{r\}$ is an {\em \dpc{r}{T}} if for every terminal pair $\{s,t\}\in T$, 
$S$ is an \sep{s}{r} or a \sep{t}{r}.\footnote{The definition of digraph pair cut used here is same as that of 
Kratsch and Wahlstr{\"{o}}m~\cite{KratschW12} where we reverse the directions of the arcs of the graph.}
\end{definition}

The problem \dpcfull\ takes as input a digraph $D$, a set of terminal pairs $T$, $r \in V(D)$ and $k \in \mathbb{N}$, and the task is to output \yes{} if and only if there is an \dpc{r}{T} in $G$ of size at most $k$.
We say that a vertex $v\in V(D)$ is {\em irrelevant} to the instance $(D,T,r,k)$
if there is no minimal \dpc{r}{T} of size at most $k$ in $D$ that contains $v$. If a vertex is not irrelevant to $(D,T,r,k)$, then we say that it is {\em relevant} to $(D,T,r,k)$.
In the following lemma, which is the main result of this subsection, we show that for an instance $(D,T,r,k)$ of \dpcfull,   
the number of in-neighbours of $r$ that belong to at least one minimal \dpc{r}{T} of size at most $k$ is upper bounded 
by $\rbound$.  In other words, we bound the number of in-neighbors of $r$ that are relevant.
\begin{lemma}
\label{lem:degreduction}
Let $(D,T,r,k)$ be an instance of \dpcfull. 
The number of vertices  in $N^{-}_D(r)$ that
are relevant to $(D,T,r,k)$ is at most $\rbound$.
Moreover, there is a deterministic algorithm that given $(D,T,r,k)$, runs in time 
$\OO(\vert T \vert \cdot n (n^{\frac{2}{3}} + m))$,
and outputs a set $R\subseteq N^{-}_D(r)$ 
of size at most $\rbound$ 
which contains all relevant vertices to $(D,T,r,k)$ in $N^{-}_D(r)$.
\footnote{In other words, 
the vertices in $N^{-}_D(r)\setminus R$ are irrelevant are irrelevant to $(D,T,r,k)$.}
\end{lemma}

Towards the proof of \Cref{lem:degreduction}, we first define which
 terminal pairs are irrelevant.



\begin{definition}
Let $(D,T,r,k)$ be an instance of \dpcfull. 
A terminal pair $\{s,t\}\in T$ is {\em irrelevant} to $(D,T,r,k)$ if any 
minimal \dpc{r}{(T\setminus \{\{s,t\}\})} in $D$ of size at most $k$ is also a minimal \dpc{r}{T} in $D$. 
\end{definition}

The following observation directly follows from the definition of irrelevant terminal pairs. 
\begin{observation}
\label{prop:irr:ter:relv}
Let $D$ be a digraph, $T$ be a set of terminal pairs, $r\in V(D)$ and $k\in {\mathbb N}$.
If $\{s,t\}\in T$ is a terminal pair irrelevant to $(D,T,r,k)$, 
then any vertex relevant to $(D,T,r,k)$ is also a vertex 
relevant to $(D, T \setminus \{\{s,t\}\},r,k)$.
\end{observation}

We now define \emph{important separators}, which have played  
an important role in the context of existing literature concerning cut related problems.
%

\begin{definition}[Important Separators, \cite{Marx06i}]
Let $D$ be a digraph. For subsets $X, Y, S \subseteq V(D)$, the set of vertices reachable from $X\setminus S$ in $D\minus S$ is denoted
by $R_D(X, S)$. An \sep{X}{Y} $S$ {\em dominates} an \sep{X}{Y} $S'$ if  $\vert S \vert \leq \vert S'\vert$ 
and $R_D(X, S')\subset R_D(X, S)$. A subset $S$ is an {\em \imsep{X}{Y}} if it is minimal, and there is no \sep{X}{Y} $S'$ that dominates $S$. 
For two vertices $s,t\in V(D)$, the term {\em \imsep{s}{t}} refers to an 
\imsep{N_D^+(s)}{N_D^-(t)} in $D\minus \{s,t\}$. For $r\in V(D)$ and $Y \subseteq V(D)$, the term 
{\em \imsep{r}{Y}} refers to an \imsep{N_D^+(r)}{Y} in $D\minus r$.
\end{definition}

\begin{lemma}[\cite{ChenLL09,Marx06i}] 
\label{lem:impsepcout}
Let $D$ be a digraph, $X, Y \subseteq V (D)$, and  $k\in {\mathbb N}$.  
The number of \imseps{X}{Y} of size at most $k$ is upper bounded by $4^k$, 
and these separators can be enumerated in time $\OO(4^k \cdot k \cdot (n+m))$.
\end{lemma}


The rest of this subsection is dedicated to the proof of \Cref{lem:degreduction}. 
That is, we design an algorithm, called $\cal{A}$, that finds a set $R$ with the properties specified by \Cref{lem:degreduction}. If $\vert N^{-}_D(r)\vert\leq \rbound$, then $N^{-}_D(r)$ 
 is the required set $R$. Thus, from now on, we assume that $\vert N^{-}_D(r)\vert> \rbound$. 
Algorithm ${\cal A}$ is an iterative algorithm. In each iteration, $\cal A$ either terminates by outputting the required set $R$, or finds an irrelevant terminal pair for the input instance, removes it 
from the set of terminal pairs, and then repeats the process. 

As a preprocessing step preceding the first call to $\cal A$, we modify the graph $D$ and the set of terminal pairs $T$ as described below. The new graph $D'$ and set of terminal pairs $T'$ would allow us to accomplish our task while simplifying some arguments in the proof.  We set $D'$ to be the digraph obtained from $D$ by adding two new vertices, $s'$ and $t'$, and two new edges, $s's$ and $t't$, for each terminal pair $\{s,t\}\in T$. 
The modification is such that if a vertex $u \in V(D)$ belonged to $\ell$ terminal pairs in $T$, then $D'$ would have $\ell$ distinct vertices corresponding to $u$. 
Now, the new set of terminal pairs is defined as $T'=\{\{s',t'\}~|~\{s,t\}\in T\}$.  
It is easy to see that any minimal \dpc{r}{T} in $D$ is also a minimal \dpc{r}{T'} in $D'$. 
Thus, to find a superset of relevant vertices for $(D,T,r,k)$ in the set $N^{-}_D(r)$, it is enough to find a superset of relevant vertices for $(D',T',r,k)$ in the set $N^{-}_{D'}(r)$.  
Therefore, from now on we can assume that our input instance is $(D',T',r,k)$, where the set $T'$ is pairwise disjoint (see \Cref{fig:D,fig:Dprime} for an illustration). Henceforth, whenever we say that a vertex is relevant (or irrelevant), we mean that it is relevant (or irrelevant) for the instance $(D',T',r,k)$.
The description of ${\cal A}$ is given in \Cref{algo:dpc}

%
%

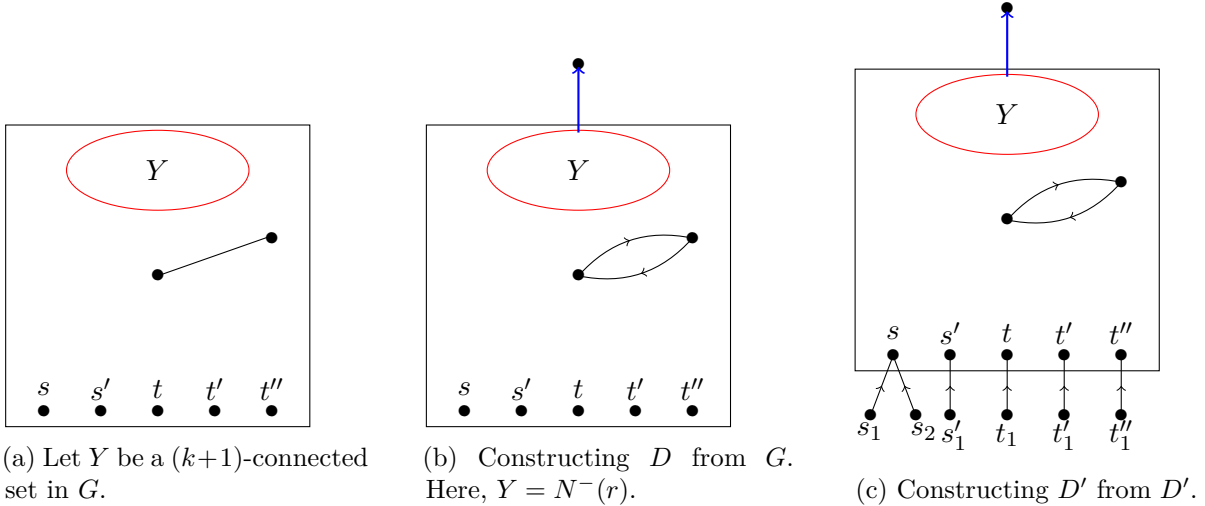
\begin{figure}[t]
\centering
\begin{subfigure}[b]{0.3\textwidth}
\begin{tikzpicture}[scale=1]
\draw (0,0) rectangle (4,4);
\draw[red] (2,3.4) ellipse (1.2cm and 0.53cm);

\node [] (a) at (0.5,0.2) {$\bullet$};
\node [] (a) at (1.25,0.2) {$\bullet$};
\node [] (a) at (2,0.2) {$\bullet$};
\node [] (a) at (2.75,0.2) {$\bullet$};
\node [] (a) at (3.5,0.2) {$\bullet$};

\node [] (a) at (0.5,0.5) {$s$};
\node [] (a) at (1.25,0.5) {$s'$};
\node [] (a) at (2,0.5) {$t$};
\node [] (a) at (2.75,0.5) {$t'$};
\node [] (a) at (3.5,0.5) {$t''$};

\node [] (a) at (2,3.4) {$Y$};

\node [] (a) at (2,2) {$\bullet$};
\node [] (a) at (3.5,2.5) {$\bullet$};
\draw (2,2) -- (3.44,2.51);

\end{tikzpicture}
                \caption{Let $Y$ be a $(k+1)$-connected set in $G$.}
            \label{fig:G}
        \end{subfigure} \hspace{5mm}
\begin{subfigure}[b]{0.3\textwidth}
\begin{tikzpicture}[scale=1]
\draw (0,0) rectangle (4,4);
\draw[red] (2,3.4) ellipse (1.2cm and 0.53cm);
\node [] (a) at (2,4.8) {$\bullet$};
\draw [->, thick, blue] (2,3.9)-- (2,4.77);

\node [] (a) at (0.5,0.2) {$\bullet$};
\node [] (a) at (1.25,0.2) {$\bullet$};
\node [] (a) at (2,0.2) {$\bullet$};
\node [] (a) at (2.75,0.2) {$\bullet$};
\node [] (a) at (3.5,0.2) {$\bullet$};

\node [] (a) at (0.5,0.5) {$s$};
\node [] (a) at (1.25,0.5) {$s'$};
\node [] (a) at (2,0.5) {$t$};
\node [] (a) at (2.75,0.5) {$t'$};
\node [] (a) at (3.5,0.5) {$t''$};

\node [] (a) at (2,3.4) {$Y$};

\node [] (a) at (2,2) {$\bullet$};
\node [] (a) at (3.5,2.5) {$\bullet$};

\draw[->-] (2,2)  to [bend left]  (3.5,2.5);
\draw[->-] (3.5,2.5)  to [bend left]  (2,2);


\end{tikzpicture}
                \caption{Constructing $D$ from $G$. Here, $Y=N^{-}(r)$.}
          \label{fig:D}
        \end{subfigure} \hspace{5mm}
\begin{subfigure}[b]{0.3\textwidth}
\begin{tikzpicture}[scale=1]
\draw (0,0) rectangle (4,4);
\draw[red] (2,3.4) ellipse (1.2cm and 0.53cm);
\node [] (a) at (2,4.8) {$\bullet$};
\draw [->, thick, blue] (2,3.9)-- (2,4.77);

\node [] (a) at (0.5,0.2) {$\bullet$};
\node [] (a) at (1.25,0.2) {$\bullet$};
\node [] (a) at (2,0.2) {$\bullet$};
\node [] (a) at (2.75,0.2) {$\bullet$};
\node [] (a) at (3.5,0.2) {$\bullet$};

\node [] (a) at (0.5,0.5) {$s$};
\node [] (a) at (1.25,0.5) {$s'$};
\node [] (a) at (2,0.5) {$t$};
\node [] (a) at (2.75,0.5) {$t'$};
\node [] (a) at (3.5,0.5) {$t''$};

\node [] (a) at (2,3.4) {$Y$};
\node [] (a) at (2,2) {$\bullet$};
\node [] (a) at (3.5,2.5) {$\bullet$};
\draw[->-] (2,2)  to [bend left]  (3.5,2.5);
\draw[->-] (3.5,2.5)  to [bend left]  (2,2);

\node [] (a) at (0.2,-0.6) {$\bullet$};
\node [] (a) at (0.8,-0.6) {$\bullet$};
\node [] (a) at (1.25,-0.6) {$\bullet$};
\node [] (a) at (2,-0.6) {$\bullet$};
\node [] (a) at (2.75,-0.6) {$\bullet$};
\node [] (a) at (3.5,-0.6) {$\bullet$};


\draw[->-] 
(0.2,-0.6) to (0.5,0.2);
\draw[->-] (0.8,-0.6) to (0.5,0.2);
\draw[->-] (1.25,-0.6) to (1.25,0.2);

\draw[->-] (2,-0.6) to (2,0.2);
\draw[->-] (2.75,-0.6) to (2.75,0.2);
\draw[->-] (3.5,-0.6) to (3.5,0.2);

\node [] (a) at (0.19,-0.8) {$s_1$};
\node [] (a) at (0.9,-0.8) {$s_2$};
\node [] (a) at (1.3,-0.85) {$s'_1$};
\node [] (a) at (2,-0.85) {$t_1$};
\node [] (a) at (2.75,-0.85) {$t'_1$};
\node [] (a) at (3.5,-0.85) {$t''_1$};

\end{tikzpicture}
                 \caption{Constructing $D'$ from $D'$.}
                \label{fig:Dprime}
        \end{subfigure}%
                 \caption{The graphs $G,D$ and $D'$ are displayed in left-to-right order, $T=\{\{s,t\},\{s,t''\},\{s',t'\}\}$ and $T'=\{\{s_1,t_1\},\{s_2,t''_1\},\{s'_1,t'_1\}\}$.}
\end{figure}

\begin{algorithm}[h]
 \If{$\vert T'\vert=0$ } {\Return{$\emptyset$}\label{step1}}  
$\widehat{T}:=\{s',t'~|~\{s',t'\}\in T'\}$. \\
Compute a minimum \sep{\widehat{T}}{r} $Z$. \label{comZ}\\
\If{$\vert Z \vert \leq \cbound$}{For each $z\in Z$, compute all \imseps{z}{r} of size at most $k$. \label{step6}\\
Mark all the vertices in $N^{-}_{D'}(r)$, which are either part of the computed important separators 
or part of $Z$.\\
 \Return{the set of marked vertices (call it $R$)} \label{outR}}
\Else{Compute a maximum set ${\cal P}$  of vertex disjoint paths from $\widehat{T}$ to $r$ (any pair of paths intersects only at $r$). \\
Let $X=V({\cal P})\cap \widehat{T}$. Let $A$ be a maximum sized subset of $X$ such that for 
any $\{s',t'\}\in T'$, $\vert A\cap \{s',t'\}\vert \leq 1$.\\
 Let $B=\{w~|~\mbox{there exists $w'\in A$ such that }\{w,w'\}\in T'\}$. 
That is, $B$ is the set of vertices that are paired with vertices of $A$ in the set of pairs $T'$.\\
Compute all \imseps{r}{B} of size at most $2k+2$ in $\overleftarrow{D'}$. \label{stepirr} \\
Mark all vertices from $B$ which are part of the computed important separators. \\
 Let $q\in B$ be an unmarked vertex and let $\{q,q'\}\in T'$. \\
$T':=T'\setminus \{\{q,q'\}\}$ and repeat from Step~\ref{step1}.}  
\caption{Input is $(G',T',r,k)$, where $T'$ is pairwise disjoint}
\label{algo:dpc}
\end{algorithm}



\begin{lemma}
\Cref{algo:dpc} outputs a set $R$ of size at most $\rbound$, which contains  all relevant vertices in $N^{-}_D(r)$.   
\end{lemma}
\begin{proof}
Notice that \Cref{algo:dpc} returns a set $R$ either in \Cref{step1} or in \Cref{outR} thus, by \Cref{lem:impsepcout}, the size of the returned set is at most $\vert Z \vert \cdot 4^k k + \vert Z \vert \leq  \rbound$.   
We now prove the correctness of the algorithm 
using induction on $\vert T'\vert$. When $\vert T'\vert=0$, then no vertex in  $N^{-}_D(r)$ is relevant 
and the algorithm returns the correct output. Now consider the induction step where $\vert T'\vert >0$. We have two cases based on 
the size of the separator $Z$ computed in \Cref{comZ}.

\paragraph*{Case 1: $\vert Z \vert \leq \cbound$.} 
In this case, Lines~\ref{step6}-\ref{outR} will be executed and \Cref{algo:dpc} will output a set $R$. We prove that $R$ contains all 
relevant vertices in $N^{-}_{D'}(r)$. Towards this, we show that if 
$S$ is a minimal \dpc{r}{T'} 
of size at most $k$ and $v\in N^{-}_{D'}(r)\cap S$,  then $v$ belongs to $R$. 
Let $S'=S\setminus \{v\}$. Since $S$ is a minimal \dpc{r}{T'}, $S'$ is not a \dpc{r}{T'}.  
Since $S$ is a  \dpc{r}{T'} and $S'$ is not a  \dpc{r}{T'}, 
there is a vertex $t\in \widehat{T}$ such that  $(i)$ $v$ is reachable from $t$ in $D'\minus S'$,
and $(ii)$ $r$ is not reachable from $t$ in $D'\minus S$.  
If $v\in Z$, then $v$ is marked and belongs to $R$. Therefore, if $v\in Z$, we are done. Thus, from now on, assume that $v\notin Z$. 
\begin{claim}
There is a vertex $z\in Z$ that belongs to  $R_{D'}(t,S)$. 
\end{claim}
\begin{proof}
From $(i)$, we have that $v\in R_{D'}(t,S')$.  
Since $Z$ is a  minimum \sep{\widehat{T}}{r}, $t\in \widehat{T}$, and $v\in R_{D'}(t,S')$, 
we have that all paths from $t$ to $v$ passes through some vertex in $Z$. 
Also, since $v\in N^{-}_{D'}(r)$ and $v\in R_{D'}(t,S')$ and $v\notin Z$, there is a vertex $z\in Z$ that belongs to $R_{D'}(t,S)$. 
\end{proof}

Let $R_t=R_{D'}(t,S)$ and $C=N^{+}_{D'}(R_t)$. Observe that $C\subseteq S$, $v\in C$ 
and $v$ is reachable from $z$ in $D'\minus (C \setminus \{v\})$. 
We claim that $C$ is a \sep{z}{r}.  If $C$ is not  a \sep{z}{r}, then there is a path from $z$ to $r$ in $D'\minus S$. 
Also, since $z\in R_{D'}(t,S)$, there is a path from $t$ to $z$ in $D'\minus S$. This implies that there is a path 
from $t$ to $r$ in $D'\minus S$ which is a contradiction to the statement $(ii)$.
Since $v$ is reachable from $z$ in $D'\minus (C \setminus \{v\})$, there is a minimal 
 \sep{z}{r} that contains $v$ and is fully contained in $C$.
Let $C'\subseteq C$ be a minimal \sep{z}{r} that contains $v$. 
Since $v\in N^{-}_{D'}(r)$ and $C'$ is a minimal \sep{z}{r}, either $C'$ is an \imsep{z}{r} or there is an \imsep{z}{r} of size at most $k$ containing $v$ which dominates $C'$. In either case,  $v$ is marked in \Cref{outR} and hence, it will be in the set $R$ 
(see \Cref{fig:cut} for an illustration).


\begin{figure}[t]
\centering
\begin{tikzpicture}[scale=1]

 \draw[red] (2,3.5) ellipse (3cm and 1.3cm);
\draw (0,0) rectangle (1,4);
\draw (5.8,0) rectangle (7.2,4);
\draw[fill=black!20] (5.8,2) rectangle (7.2,4);
\node [] (a) at (0.5,3.5) {$\bullet$};
\node [] (a) at (6.5,3.5) {$\bullet$};
\node [] (a) at (9.5,2) {$\bullet$};

\node [] (a) at (0.5,3.78) {$t$};
\node [] (a) at (6.5,3.7) {$v$};
\node [] (a) at (9.5,1.7) {$r$};

\node [] (a) at (0.5,-0.5) {$\widehat{T}$};
\node [] (a) at (6.5,-0.5) {$S$};

\node [] (a) at (3.5,3.5) {$R_t=R_{D'}(t,S)$};

\node [] (a) at (6.5,2.3) {$N^+(R_t)$};

%

\draw [->] (6.5,3.5)-- (9.42,2.05);

\draw [->, thick] (5,3.6)-- (6.2,3);

\end{tikzpicture}
 \caption{Here, the ellipse contains the set of vertices reachable from $t$ in $D'\minus S$, denoted by $R_t$. 
The rectangle colored grey represents $N^+(R_t)$ which includes $v$}\label{fig:cut}
\end{figure}
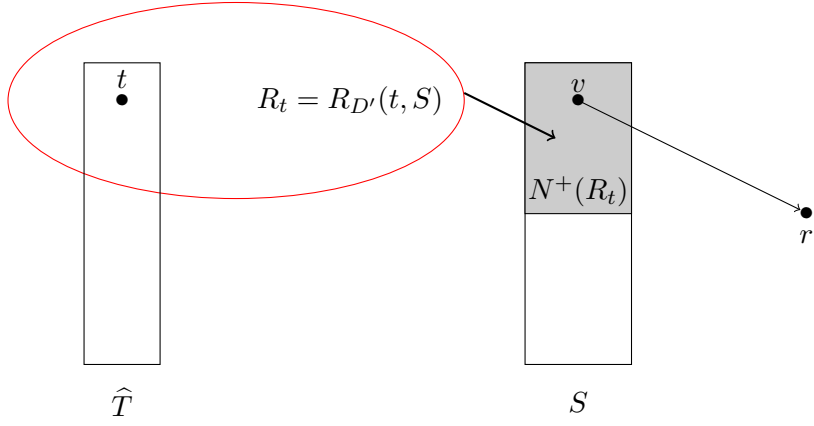

\paragraph*{Case 2: $\vert Z \vert > \cbound$.} 
In this case, we prove that there, indeed, exists an unmarked vertex $q\in B$ and the pair $\{q,q'\}$ is an irrelevant terminal pair. 
Notice that in \Cref{stepirr}, we have computed all \imseps{r}{B} of size at most $2k+2$ for some $B$. 
By \Cref{lem:impsepcout}, the total number of vertices in all these separators together is at most $16^k\cdot 32(k+1)$. So we should have marked 
at most $16^k\cdot 32(k+1)$ vertices in $B$. We first claim that $\vert B\vert >16^k\cdot 32(k+1)$, which ensures the existence of 
an unmarked vertex in $B$.  By the definition of $A$, the size of $A$ is at least  $\vert Z \vert /2>16^k\cdot 32(k+1)$, 
because there are $\vert Z \vert$ vertex disjoint paths from $\widehat{T}$ to $r$, only intersecting at $r$. 
By the definition of $B$, $\vert B \vert =\vert A \vert >16^k\cdot 32(k+1)$. Since we proved that we have only 
marked at most $16^k\cdot 32(k+1)$ vertices in $B$, this implies that there is an unmarked vertex $q$ in $B$. 
Let $q'$ be the unique  vertex such that $\{q,q'\}\in T'$ (such a unique vertex exists because $T'$ is pairwise disjoint). 

Now we show that $\{q,q'\}$ is an irrelevant terminal pair. 
Let $S$ be a minimal \dpc{r}{(T'\setminus \{\{q,q'\}\})} of size at most $k$. We need to show that $S$ is also 
a \dpc{r}{T'}. 
 %
%
%
We know that there are $\vert Z\vert$ vertex disjoint paths ${\cal P}$ from $\widehat{T}$ to $r$, where the paths intersect only at $r$. 
Since  $Z$ is a minimum  \sep{\widehat{T}}{r}, $\vert \widehat{T}\vert\geq \vert Z\vert$. 
Recall the definition of $A$ and $B$ from the description of the algorithm. 
Let $A_r$ be the set of vertices 
in $A\setminus \{q'\}$ such that $r$ is reachable from each vertex in $A_r$ in $D'\minus S$, that is, $A_r=\{u\in A\setminus \{q'\}~|~r\in R_{D'}(u,S)\}$. 
Let $B_{r}$ is the 
set of vertices in $B$ such that $r$ is reachable from any vertex in $B$ in $D'\minus S$, that is, $B_{r}=\{u'\in B~|~r\in R_{D'}(u',S)\}$. Since there are $\vert A\vert$ vertex disjoint paths from $A$ 
to $r$ (which intersect only at $r$) and $\vert S \vert \leq k$ $\vert A_r\vert \geq \vert A\vert -(k+1)$. 
Since $S$ is an \dpc{r}{(T'\setminus \{\{q,q'\}\})}, the vertices in $B$ which are paired with a vertex in $A_r$ are not reachable from 
$r$ in $\overleftarrow{D'}\minus S$. This implies that $\vert B_{r}\vert \leq k+1$.
Let $Q=S\cup B_r \cup \{q\}$. Notice that $q\in Q$ and $Q$ is a \sep{r}{B} in $\overleftarrow{D'}$ of size at most $2k+2$. 
If $q$ is not reachable from $r$ in $\overleftarrow{D'}\minus S$, then $S$ is, indeed, a 
\dpc{r}{T'}, because $S$ is a \dpc{r}{(T'\setminus \{\{q,q'\}\})}. 
In what follows we show that it is always the case, that is, $q$ is not reachable from $r$ in $\overleftarrow{D'}\minus S$. 
Suppose not. 
Since $q$ is reachable  from $r$ in $\overleftarrow{D'}\minus S$
and all the vertices in $Q \setminus S$ have no out-neighbours in $\overleftarrow{D'}$ (by construction of $D'$), any path from 
$r$ to $q$ in  $\overleftarrow{D'}\minus S$ will not contain any vertex from $Q\setminus \{q\}$. 
This implies that there is a minimal \sep{r}{B} $Q'\subseteq Q$ containing $q$. 
Hence, either $Q'$ is an \imsep{r}{B} of size at most $2k+2$ or all the 
\imseps{r}{B} which dominate $Q'$ will contain $q$. 
This is implies that $q$ is marked, which is a contradiction.  

Thus, we have shown that in this case there is an irrelevant terminal pair $\{q,q'\}\in T'$, and by Observation\Cref{prop:irr:ter:relv} 
and induction hypothesis, \Cref{algo:dpc} will output the required set.   
\end{proof}
\begin{lemma}
\Cref{algo:dpc} runs in time $\OO(\vert T' \vert \cdot n( n^{\frac{2}{3}}+ m))$.
\end{lemma}
\begin{proof}
The number of times each step of the algorithm will get executed is at most $\vert T'\vert$. By \Cref{prop:ststime}, \Cref{comZ} takes time 
$\OO(mn)$.
By \Cref{lem:impsepcout}, the time required to enumerate important separators in \Cref{step6,stepirr} 
is bounded by $\OO(4^{2k} \cdot k \cdot (n+m))$. The time required compute ${\cal P}$ in \Cref{stepirr} is $\OO(mn)$
by \Cref{prop:ststime}. Thus, the total running time of \Cref{algo:dpc} is $\OO( \vert T' \vert (mn + 4^{2k} \cdot k \cdot (n+m)))$.
Recall that, we could safely assume that
$|V(D')|=n > \rbound$. Since, $n > \rbound$, $4^{2k} \cdot k < n^{\frac{2}{3}}$.
Hence, the claimed running time of the algorithm follows. 
\end{proof}
\newcommand{\dgbound}{64^{k+1}\cdot 2k^2}
\newcommand{\cdbound}{64^{k+1} (k+1)^2}
\newcommand{\cdboundw}{64^{k} \cdot k^2}
\newcommand{\topfree}{64^{k+2} \cdot 2\cdot (k+1)^2}
\newcommand{\topfreed}{64^{k+2} \cdot 4\cdot (k+1)^2}

\subsection{Covering Small Multicuts in a Subgraph without Highly Connected Set}
\label{subsect:multtool}

In this section, we prove that given a graph $G$, a set of terminal pairs $T=\{\{s_1,t_1\},\ldots, \{s_{\ell},t_{\ell}\}\}$ and an integer 
$k$, there is a polynomial time algorithm which finds a pair $(G^{\star}, T^\star)$, where $G^{\star}$ is an induced subgraph of $G$ such that it has no $(k+2)$-connected sets of size $2^{\OO(k)}$ and $T^\star \subseteq T$ such that for any $S \subseteq V(G)$ of size at most $k$, $S$ is a minimal multicut of $T$ in $G$ if and only $S$ is a subset of $V(G^\star)$ and $S$ is a minimal multicut of $T^{\star}$ in $G^\star$. This statement is formalized in Lemma~\ref{lem:main:top}.
%
Before stating Lemma~\ref{lem:main:top}, we give definitions of a $k$-connected set in a graph $G$ and a $k$-connected graph. 


\begin{definition}[$k$-connected set and graphs]
For any $k \in \mathbb{N}$ and a graph $G$, a subset $Y$ of the vertices of $G$ is called a \emph{$k$-connected set} in $G$, 
if for any $u,v\in Y$ there are at least $k$ internally vertex disjoint paths 
from $u$ to $v$ in $G$. 
The graph $G$ is called a \emph{$k$-connected graph} if $V(G)$ is a $k$-connected set in $G$. Equivalently, the graph $G$ 
is $k$-connected, if the size of a mincut in $G$ is at least $k$. 
\end{definition}


\begin{lemma}[Degeneracy Reduction Lemma]
\label{lem:main:top}
Let $G$ be a graph, $T$ be a set of terminal pairs and $k\in{\mathbb N}$. Let $C$ be the set of 
all minimal multicuts of $T$ of size at most $k$ in $G$. 
There is a deterministic 
algorithm which runs in time 
$\OO(\vert T\vert \cdot n^2 (n^{\frac{2}{3}} +m ) + k n^3 (n+m))$
and outputs an induced subgraph  $G^\star$ of $G$ and a subset $T^\star\subseteq T$ 
such that  
\begin{enumerate}
\item 
for any $S\subseteq V(G)$ with $\vert S\vert \leq k$, 
$S$ is a minimal multicut of $T$ in $G$  if and only if $S\subseteq V(G^\star)$ and $S$ is a minimal multicut of $T^\star$ in $G^\star$, and 
\item there is no $(k+2)$-connected set of size at least ${64}^{k+2} \cdot 4 {(k+2)}^2$ in $G^\star$. 
\end{enumerate}
\end{lemma}




The proof of Lemma~\ref{lem:main:top} requires some auxiliary lemmas which we discuss below.
Recall the definition of the problem \mc\ from \Cref{sec:AppImulticut}. Let $(G,T,k)$ be an instance of \mc.
We say that a vertex $v\in V(G)$ is \emph{irrelevant} to $(G,T,k)$
if no minimal multicut of $G$ 
of size at most $k$ in $G$ contains $v$.  
Lemma~\ref{lem:condpc} states that if a graph has a sufficiently large $(k+2)$-connected set, then many of its vertices are irrelevant to the given \mc\ instance. Such a statement is deduced by establishing a relation between the multicuts of the given instance and the digraph pair cuts of practically the same instance. This relationship then relates the irrelevant vertices to the instance of \mc\ with the irrelevant vertices to the instance for \dpcfull. 


\begin{lemma}
\label{lem:condpc}
Let $G$ be a graph, $T$ be a set of terminal pairs, $k\in {\mathbb N}$ and 
$Y$ be a $(k+1)$-connected set in $G$.  
Let $D$ be a digraph obtained by 
adding a new vertex $r$, whose in-neighbours are the vertices of $Y$, and replacing each edge of $G$ by two arc, with the same endpoints,
in opposite orientations. Any irrelevant vertex to the instance $(D,T,r,k)$ of \dpcfull{}
is also an irrelevant vertex to the instance $(G,T,k)$ of \mc.  
\end{lemma}

\begin{proof}
The construction of $D$ from $G$ is illustrated in \Cref{fig:G,fig:D}. 
To prove the lemma it is enough to show that any multicut of $T$ of size at most $k$ in $G$ is an \dpc{r}{T} in $D$.
Let $C$ be a multicut of $T$ of size at most $k$ in $G$.  
We claim that $C$ is an \dpc{r}{T} in $D$. Suppose not. Then, there is a pair  $\{s,t\}\in T$ such that there is a path from $s$ to $r$ and $t$ to $r$ in $D \minus C$. Since the in-neighbors of $r$ are the vertices of $Y$, there exist $u_1,u_2 \in Y$, $u_1$ may be equal to $u_2$, such that there are two paths, one from $s$ to $u_1$ and other from 
$t$ to $u_2$,in $G\minus C$. 
If $u_1=u_2$, then $s$ and $t$ are in the same connected component 
of $G\minus C$, which is a contradiction. 
Otherwise, since $Y$ is a $(k+1)$-connected set in $G$ and $u_1, u_2 \in Y$, there are $k+1$ internally vertex disjoint paths from 
$u_1$ to $u_2$. Since $\vert C\vert\leq k$, there exists a path between $u_1$ and $u_2$ in $G\minus C$, and hence a path between $s$ and $t$ in $G\minus C$, which is a contradiction.  
\end{proof}

Using \Cref{lem:condpc,lem:degreduction}, one can find irrelevant vertices to the given instance of \mc, if the graph in the instance has a $(k+1)$-connected set $Y$ of size strictly more than $\rbound$ and the set $Y$ is explicitly given as input.
So the next task is to design an algorithm that finds a $(k+1)$-connected set in a graph of a given size, if it exists. This algorithm comes from \Cref{lem:largetopminor}.

\begin{lemma}
\label{lem:largetopminor}
There is an algorithm 
which given a graph $G$ and $k,d\in {\mathbb N}$, $k\leq d$, runs in time $\OO(k\cdot n^{2}(n+m))$, and 
either concludes that there is no $k$-connected set of size at least $4d$ in $G$ or 
outputs a $k$-connected 
set in $G$ of size at least $d+1$.
\end{lemma}

The proof of \Cref{lem:largetopminor} requires an auxiliary lemma (\Cref{lem:dconnalgo}) which we prove next. \Cref{lem:dconnalgo} is an algorithmic verison of the 
following famous result of Mader~\cite{Mader1972} which says that 
if a graph has large average degree (or degeneracy), then 
it contains a $(d+1)$-connected subgraph. 

\begin{lemma}[\cite{Mader1972}]
\label{lem:mader}
Let $d\in {\mathbb N}\setminus \{0\}$. Every graph $G$ with average degree at least $4d$ has a $(d+1)$-connected subgraph. 
\end{lemma} 

The proof of \Cref{lem:mader} given in \cite{diestel,Sudakov16} can be modified to get a polynomial time algorithm.  
The following lemma, 
an algorithmic version of \Cref{lem:mader}, is written in terms of the degeneracy of the graph. 

\begin{lemma}
\label{lem:dconnalgo}
There is an algorithm which, for any $d\in {\mathbb N}\setminus \{0\}$, given a graph $G$ with degeneracy at least $4d$, runs 
in time $\OO(mn+n^2\log n)$, and outputs a $(d+1)$-connected subgraph of $G$. 
\end{lemma}

\begin{proof}
The algorithm first constructs a subgraph $H$ of $G$ which has minimum degree at least $4d$.
To do so, first set $H:=G$. If the minimum degree of $H$ is at least $4d$, then we are done. 
Otherwise, let $v$ be a vertex of $H$ of degree at most $4d-1$. Set $H:=H \minus v$ and repeat this process.  
Since the degeneracy of $G$ is at least $4d$, the procedure will end up in a subgraph of $G$ that has minimum degree at least $4d$. 
The naive implementation of the above procedure takes time $\OO(mn)$. 
\begin{claim}\label{claim:mindegbounds}
For any $d \in \mathbb{N} \minus \{0\}$, if the minimum degree of a graph $H$ is at least $4d$, then $\vert V(H) \vert \geq 2d+1$ and $\vert E(H) \vert \geq 2d(\vert V(H)\vert -d -\frac{1}{2})$.
\end{claim}
\begin{proof}
Since minimum degree of $H$ is at least $4d$, clearly $ \vert V(H) \vert \geq 4d+1 \geq 2d+1$. Also, since $\sum_{v \in V(H)} {deg}_G(v) = 2 \vert E(H) \vert$ and for all $v \in V(H)$ ${deg}_G(v) \geq 4d$, $\vert E(H) \vert \geq 2d \vert V(H) \vert \geq 2d(|V(H)| -d -\frac{1}{2})$. 
\end{proof}
From \Cref{claim:mindegbounds}, we conclude that $\vert V(H) \vert \geq 2d+1$ and $\vert E(H) \vert \geq 2d(\vert V(H)\vert -d -\frac{1}{2})$. 
Thus, from the following claim (\Cref{claim:dconn}), one can infer that $H$ has a $(d+1)$-connected subgraph. Using this claim, we will later give an algorithm that actually computes a $(d+1)$-connected subgraph of $H$, whose correctness will follow from the proof of \Cref{claim:dconn}. 

\begin{claim}
\label{claim:dconn}
Let $H$ be any graph and $d\in {\mathbb N}\setminus \{0\}$ such that $\vert V(H)\vert \geq 2d+1$ and $\vert E(H) \vert \geq 2d(\vert V(H)\vert -d -\frac{1}{2})$. Then $H$ has a $(d+1)$-connected subgraph.
\end{claim}
\begin{proof}
We prove the claim using induction on $\vert V(H)\vert$. 
The base case of the induction is when 
$\vert V(H)\vert = 2d+1$. From the premises of the claim, if $\vert V(H)\vert = 2d+1$, $\vert E(H)\vert\geq 2d(2d+1-d-\frac{1}{2})=2d(d+\frac{1}{2})=\binom{2d+1}{2}$. Since a graph on $2d+1$ vertices can have at most $\binom{2d+1}{2}$ edges, $H$ is a clique on ${2d+1}$ vertices, which is a $(d+1)$-connected graph. 
Now consider the induction step where 
$\vert V(H)\vert > 2d+1$. Suppose there is a vertex $v\in V(H)$ such that $deg_H(v)\leq 2d$. Then $\vert V(H \minus v) \vert \geq 2d+1$ and $ \vert E(H \minus v) \vert \geq  \vert E(H) \vert - 2d \geq 2 d (\vert V(H-v) \vert -d - \frac{1}{2})$. Thus, from the induction hypothesis, there is a $(d+1)$-connected subgraph in $H \minus v$. From now on, we can assume that the degree of each vertex in $H$ is at least $2d +1$. Suppose $H$ itself is a $(d+1)$-connected graph, then we are done. If not, then there exists a mincut, say $Z$, of $H$, of size at most $2d$. 
Let $U_1\uplus U_2$ be a partition of $V(G)\setminus Z$ 
such that there is no edge between a vertex in $U_1$ and a vertex in $U_2$, and $U_1 , U_2 \neq \emptyset$. 
Let $A=Z\cup U_1$ and $B=Z\cup U_2$. 
We claim that either $H[A]$ or $H[B]$ satisfy the premises of the claim. 
Notice that all the neighbors of any vertex 
$s\in U_1$ are in $A$ and all the neighbors of any vertex $t\in U_2$ are in $B$.  Also since, $deg_H(s),deg_H(t)\geq 2d+1$, we have that 
$\vert A \vert \geq 2d+1$ and $\vert B \vert \geq 2d+1$. Thus,the vertex set cardinality constraint stated in the premise of the claim is met for 
both $H[A]$ and $H[B]$. Suppose that, the edge set cardinality constraint stated in the premise of the claim is not met for both 
both $H[A]$ and $H[B]$. Then we have the following.
\begin{eqnarray*}
\vert E(H) \vert &\leq& \vert E(H[A]) \vert + \vert E(H[B]) \vert \\
&<& 2d(\vert A \vert -d-\frac{1}{2})+2d(\vert B \vert -d-\frac{1}{2}) \\
&=&2d(\vert A \vert+\vert B \vert -2d-1)\\
&\leq&2d(\vert V(H) \vert +d -2d-1)\\
&<&2d(\vert V(H) \vert -d-\frac{1}{2}). 
\end{eqnarray*}
This is a contradiction to the fact that $\vert E(H) \vert \geq 2d(\vert V(H) \vert -d-\frac{1}{2})$. Therefore, either $H[A]$ 
or $H[B]$ satisfy the premises of the claim. Moreover, notice 
that $\vert A \vert < \vert  V(H)\vert$ and  $\vert B \vert < \vert  V(H)\vert$, because $U_1, U_2 \neq \emptyset$. 
Thus, by the induction hypothesis the claim follows. 
\end{proof}

The above proof can easily be turned in to an algorithm. This is explained below.
Our algorithm for finding a $(d+1)$-connected subgraph of $H$ works as follows. 
It first tests whether $H$ itself is a $(d+1)$-connected graph - this can be done by computing a mincut of $H$ (using the algorithm of Proposition~\ref{prop:mincuttime}) and then testing whether the size of a mincut of $H$ is at least $d+1$. 
If $H$ is a $(d+1)$-connected graph, then our algorithm outputs $H$. Otherwise, if there 
is a vertex of degree at most $2d$ in $H$, then it recursively finds a $(d+1)$-connected subgraph in 
$H\minus v$. If all the vertices in $H$ have degree at least $2d+1$, then it finds a mincut 
$Z$ in $H$ (using the algorithm of Propositon~\ref{prop:mincuttime}). It then constructs vertex sets $A$ and $B$ as mentioned in the proof of \Cref{claim:dconn}. 
It is proved in \Cref{claim:dconn} that either $H[A]$ or $H[B]$ satisfy the premises of \Cref{claim:dconn}, 
and it can be tested in linear time whether a graph satisfies the premises of \Cref{claim:dconn}. If $H[A]$ satisfies the premises of \Cref{claim:dconn}, then our algorithm 
recursively finds a $(d+1)$-connected subgraph in $H[A]$. Otherwise, our algorithm recursively find a $(d+1)$-connected subgraph in $H[B]$. 

Note that this algorithm makes at most $n$ recursive calls and in each recursive call 
it runs the algorithm of \Cref{prop:mincuttime} and does some linear time testing. Thus, given a graph $H$ of minimum degree at least $2d$, this algorithm runs in time $\OO(n (m + n \log n) )$ and outputs a $(d+1)$-connected subgraph of $H$.
The algorithm claimed in the lemma first constructs a subgraph $H$ of $G$ of minimum degree at least $2d$, as described earlier, in time $\OO(mn)$ and takes additional $\OO(mn+ n^2 \log n)$ time to output a $(d+1)$-connected subgraph of $H$. Thus, the total running time of this algorithm is $\OO(mn+n^2\log n)$.  
\end{proof}





We are now equipped to give the proof of \Cref{lem:largetopminor}. 

\begin{proof}[Proof of \Cref{lem:largetopminor}]
The algorithm first constructs an auxiliary graph $G^*$ as follows. The vertex set of $G^*$ is $V(G)$ and for any $u,v\in V(G^*)$,$uv \in E(G^*)$ if and only if the size of a minimum \sep{u}{v} in $G$ is at least $k$ (that is, 
there are at least $k$ internally vertex disjoint paths from $u$ to $v$ in $G$). 
It then checks whether the degeneracy of $G^*$ is at least $4d-1$ or not.  
If the degeneracy of $G^*$ is strictly less than $4d-1$, then the algorithm outputs that 
there is no $k$-connected set in $G$ of size at least $4d$. 
Otherwise, the degeneracy of $G^*$ is at least $4d-1 \geq 4(d-1)$. In this case, the algorithm applies the algorithm of \Cref{lem:dconnalgo} for $(G^*,d-1)$, which outputs
a $d$-connected subgraph $H$ of $G^*$. Since $H$ is a $d$-connected subgraph, $\vert V(H) \vert \geq d+1$. Since, $k \leq d$, $H$ is $k$-connected in $G^*$. 
The algorithm outputs $V(H)$ as the $k$-connected set in $G$. 

To prove the correctness of the algorithm, we need to prove the following two statements.
\begin{enumerate}
 \item When our algorithm reports that there is no $k$-connected set in $G$ of size at least $4d$, that is, when degeneracy of $G^*$ is at most $4d-2$, then the graph $G$ has no $k$-connected set of size at least $4d$. 
 \item When our algorithm outputs a set, that is, when degeneracy of $G^*$ is at least $4d-1$, then the set outputted is a $k$-connected set in $G$ of size at least $d+1$. In other words, if degeneracy of $G^*$ is at least $4d-1$, then the set $V(H^*)$ is $k$-connected in $G$ and has size at least $d+1$.
\end{enumerate}

For the proof of the first statement, observe that when $G$ has a $k$-connected set, say $Y$, of size at least $4d$, then $G^*[Y]$ is a clique. Hence, the degeneracy of $G^*$ is at least $4d-1$. 
For the proof of the second second, we have already argued that the size of $V(H)$ is at least $d+1$ and that $H$ is a $k$-connected subgraph in $G^*$. We will now prove $V(H)$ is a $k$-connected set in $G$.

\begin{claim}
$V(H)$ is a $k$-connected set in $G$. 
\end{claim} 
\begin{proof}
Observe that, it is enough to show that for any $u,v\in V(H)$ and any $C \subseteq V(G)\setminus \{u,v\}$ of size 
strictly less than $k$, there is a path from $u$ to $v$ in $G\minus C$.
Since $H$ is a $k$-connected subgraph of $G^*$, there is a path from 
$u$ to $v$ in $G^*\minus C$. 
Let $w_1w_2\ldots w_{\ell}$, where $w_1=u$ and $v=w_{\ell}$, be a path from $w_1$ to $w_{\ell}$ in  
$G^*\minus C$.
Since for any $i\in [\ell-1]$, $w_iw_{i+1}\in E(G^*)$, there are at least $k$ vertex 
disjoint paths from  $w_i$ to $w_{i+1}$ in $G$. Also, since $\vert C\vert < k $, 
there is a path from $w_i$ to $w_{i+1}$ in $G\minus C$. This implies 
that there is a path from $w_1=u$ to $w_{\ell}=v$ in $G\minus C$, proving that $H$ is a $k$-connected set in $G$.  
\end{proof}
This finishes the proof of correctness of our algorithm.
We now analyse the total running time of the algorithm.
The graph $G^*$ can be constructed in time $\OO(k \cdot n^{2}(n+m))$ using \Cref{prop:ststime}. Also, checking whether the graph has degeneracy at least $4d-1$ can be done in time $\OO(mn)$. 
Since $G^*$ could potentially have $\OO(n^{2})$ edges, by \Cref{lem:dconnalgo}, the subgraph $H$ can be computed in time $\OO(n^{3})$. Thus the total running time of our algorithm is $\OO(k \cdot n^{2}(n+m))$.  
\end{proof}

\begin{lemma}
\label{lem:irrmctopminor}
There is an algorithm that given a graph $G$, a set of terminal pairs $T$ and $k\in {\mathbb N}$, runs in time
$\OO(\vert T\vert \cdot n (n^{\frac{2}{3}} +m ) + k n^2 (n+m))$
and, 
either correctly concludes that $G$ does not contain a $(k+1)$-connected set of size at least ${64}^{k+1} \cdot 4(k+1)^2$ or 
finds an irrelevant vertex for the instance $(G,T,k)$ of \mc.
\end{lemma}
\begin{proof}
Let $d = \cdbound$.
Our algorithm first runs the algorithm of \Cref{lem:largetopminor}  
on the instance $(G,k+1,d)$. 
If this algorithm (of \Cref{lem:largetopminor}) concludes 
that there is no $(k+1)$-connected set in $G$ of size at least $4d$, 
then our algorithm returns the same. 
Otherwise, the algorithm of \Cref{lem:largetopminor} outputs  
a $(k+1)$-connected set $Y$ in $G$ of size at least $d+1$. 
Our algorithm then creates 
a digraph $D$ as mentioned in \Cref{lem:condpc}. It then applies the algorithm of \Cref{lem:degreduction} 
and compute a set $Z$ of irrelevant vertices for the instance $(D,T,r,k )$ of \dpcfull\ in the set $Y$. From \Cref{lem:condpc}, $Z$ is also a set of irrelevant vertices for the instance $(G,T,k)$ of \mc. Since $|Y| \geq d+1$ and the number of relevant vertices for $(D,T,r,k)$ in the set $Y$ is at most $d$ (from \Cref{lem:degreduction}), $Z \neq \emptyset$. 
Our algorithm then outputs an arbitrary vertex $v$ from the set $Z$ as an irrelevant vertex for $(G,T,k)$. 


By \Cref{lem:largetopminor,lem:degreduction},  
the total running time of our algorithm is 
$\OO(\vert T\vert \cdot n (n^{\frac{2}{3}} +m ) + k n^2 (n+m))$.
\end{proof}

\begin{lemma}
\label{lem:main:toponedel}
There is an 
algorithm which given as input a graph $G$, a set of terminal pairs $T$ and $k\in {\mathbb N}$, runs in time  
$\OO(\vert T\vert \cdot n (n^{\frac{2}{3}} +m ) + k n^2 (n+m))$
and, either concludes that 
there is no $(k+2)$-connected set of size at least ${64}^{k+2} \cdot 4{(k+2)}^2$ in $G$, 
or outputs a vertex $v\in V(G)$ such that   
for any $S\subseteq V(G)$ with  $\vert S\vert \leq k$, 
$S$ is a minimal multicut of $T$ in $G$  if and only if $S\subseteq V(G)\setminus \{v\}$ and $S$ is a minimal multicut of 
$T'=\{\{s,t\}\in T~|~v \notin \{s,t\}\}$ in $G\minus v$.  
\end{lemma}

\begin{proof}
This algorithm runs the algorithm of \Cref{lem:irrmctopminor} on the instance $(G,T,k+1)$.
If the algorithm of \Cref{lem:irrmctopminor}  
outputs that 
there is no $(k+2)$-connected set of size ${64}^{k+2} \cdot 4 {(k+2)}^2$ in $G$, then our algorithm reports the same.
Otherwise, let $v$ be the vertex retuned by the algorithm of \Cref{lem:irrmctopminor}, which is irrelevant for $(G,T,k+1)$ (from \Cref{lem:irrmctopminor}), then it also returns $v$.
The running time of our algorithm follows from \Cref{lem:irrmctopminor}. 

We now prove the correctness of this algorithm. For the forward direction, $S \subseteq V(G)$ such that  $\vert S \vert \leq k$ and $S$ is a minimal 
multicut of $T$ in $G$. By the definition of an irrelevant vertex for the instance $(G,T,k+1)$, we conclude that $S \subseteq V(G) \setminus \{v\}$. 
Since $T' \subseteq T$ and $G$ is a supergraph of $G \minus v$, $S$ is a multicut of $T'$ in $G \minus v$. 
Suppose, for the sake of contradiction, that $S$ is not a minimal multicut of $T'$ in $G \minus v$. Then, there exists $S' \subset S$ such that $S'$ is a minimal multicut of $T'$ in $G \minus v$. If $S'$ is multicut of $T$ in $G$, then we contradict the fact that $S$ is a minimal multicut of $T$ in $G$. Otherwise, there exists $S'' \subseteq S \cup \{v\}$ and $v \in S''$, such that $S''$ is a minimal multicut of $T$ in $G$, which contradicts that $v$ is an irrelevant vertex for $(G,T,k+1)$. 
Hence we have proved that $S$ is a minimal 
multicut of $T'$ in $G\minus v$. 

For the backward direction, let $S\subseteq V(G) \setminus \{v\}$ such that $S$ is a minimal 
multicut of $T'$ in $G \minus v$. If $S$ is a multicut of $T$ in $G$, then $S$ has to a minimal multicut of $T$ in $G$ else it would contradict that $S$ is a minimal multicut of $T'$ in $G \minus v$.
Otherwise, $S \cup \{v\}$ is a multicut of $T$ in $G$, because all the terminal pairs in $T\setminus T'$ contains $v$. 
Let $S'\subseteq S\cup \{v\}$ be a minimal multicut of $T$ in $G$. Note that $v \in S'$ and $\vert S' \vert \leq k+1$. This contradicts the fact that $v$ is an irrelevant vertex for $(G,T,k+1)$.

\end{proof}

\Cref{lem:main:top} can easily be proved by applying \Cref{lem:main:toponedel} at most $n$ times. 

\paragraph*{Stable Multicut on General Graphs.} 
With the power of our Indepedent Set Covering Lemmas ( Lemmas~\ref{lemma:riscl}, ~\ref{lemma:discl} and ~\ref{lemma:discl2}) and the Degeneracy Reduction Lemma (\Cref{lem:main:top}) in hand, we are now ready we prove that \imc\ is \FPT. 
Towards that, we first prove the following lemma which establishes a relationship between the degeneracy of the graph and the $k$-connected sets in the graph.  
\begin{lemma}
\label{lem:condeg}
Let $k,d\in {\mathbb N}$ such that $k\leq d+1$. Let $G$ be a graph which does not contain 
a $k$-connected set of size at least $d$. Then, the degeneracy of $G$ is at most $4d-1$. 
\end{lemma}
\begin{proof}
For the sake of contradiction, assume that the degeneracy of $G$ is at least $4d$. 
Then, by \Cref{lem:dconnalgo}, there is a $(d+1)$-connected subgraph $H$ of $G$. 
Since $k \leq d+1$ and $\vert V(H) \vert \geq d+2$, we have that $V(H)$ is $k$-connected 
set in $G$ of size at least $d+2$, which is a contradiction.    
\end{proof}

\begin{theorem}
\imc\ can be solved in time $2^{\OO(k^3)} \cdot n^3(n+m)$. 
\end{theorem}
\begin{proof}
Let $(G,k)$ be an instance of \imc. First, we apply \Cref{lem:main:top} and get an equivalent 
instance $(G^*,T^*)$, where $G^*$ does not contain any $(k+2)$-connected set of size ${64}^{k+2} \cdot 4 {(k+2)}^2$. 
Then, by \Cref{lem:condeg}, the degeneracy of $G^*$ is at most ${64}^{k+2} \cdot 16 {(k+2)}^2 - 1$. Now, we apply 
\Cref{thm:imc} and get the solution. 
The running time of the algorithm follows from \Cref{lem:main:top,thm:imc}. 
\end{proof}


\section{Conclusion}\label{sec:conclusion}

In this paper we presented two new combinatorial tools for the design of parameterized algorithms. The first was a simple linear time randomized algorithm that given as input a $d$-degenerate graph $G$ and integer $k$, outputs an independent set $Y$, such that for every independent set $X$ in $G$ of size at most $k$ the probability that $X$ is a subset of $Y$ is at least $\left({(d+1)k \choose k} \cdot k(d+1)\right)^{-1}$. We also introduced  the notion of a $k$-{\em independence covering family} of a graph $G$.  The second tool was a new (deterministic) polynomial time graph sparsification procedure that given a graph $G$, a set $T = \{\{s_1, t_1\}, \{s_2, t_2\}, \ldots, \{s_\ell, t_\ell\}\}$ of terminal pairs, and an integer $k$ returns an induced subgraph $G^\star$ of $G$ that maintains {\em all} of the inclusion minimal multi-cuts of $G$ of size at most $k$, and does not contain any $(k+2)$-vertex connected set of size $2^{\OO(k)}$. Our new tools yielded new \FPT algorithms for {\sc Stable} $s$-$t$ {\sc Separator}, {\sc Stable Odd Cycle Transversal}, and {\sc Stable Multicut} on general graphs, and for {\sc Stable Directed Feedback Vertex Set} on $d$-degenerate graphs, resolving two problems left open by Marx et al.~\cite{marx2013finding}. One of the most natural direction to pursue further is to find more applications of our tools than given in the paper. Apart from this there are several natural questions that arise form our work. 

\begin{enumerate}
\setlength{\itemsep}{-2pt}
\item In the {\sc Stable Multicut} problem we ask for a multicut that forms an independent set. 
Instead of requiring that  the solution $S$ is independent, we could require that it induces a graph that belongs to a hereditary graph class $\cal G$. Thus, corresponding to each hereditary graph class $\cal G$, we get the problem 
$\cal G$-{\sc Multicut}. Is $\cal G$-{\sc Multicut} \FPT? Concretely, let $\mathscr S$ be the set of forests then is 
$\mathscr S$-{\sc Multicut} \FPT? 
\item Given a hereditary graph class $\cal G$, we can define the notion of  $k$-$\cal G$ covering family, similar to 
 $k$-independence covering family. Does there exist other hereditary families,  apart from the family of independence sets,  such that  $k$-$\cal G$ covering family of \FPT size exists?
\item Observe that for all the problems whose non-stable version admit $2^{\cO(k)} n^{\cO(1)}$ time algorithm on general graphs, such as $s$-$t$ {\sc Separator} and {\sc Odd Cycle Transversal}, we get $2^{\cO(k)} n^{\cO(1)}$ time algorithm  for these problems on graphs of bounded degeneracy.  As a corollary, we get  $2^{\cO(k)} n^{\cO(1)}$ time algorithm  for these problems on planar graphs, graphs excluding some fixed graph $H$ as minor or a topological minor and  
graphs of bounded degree. A natural question is whether these problems admit $2^{\cO(k)} n^{\cO(1)}$ time algorithm on general graphs. 
\end{enumerate}

\bibliographystyle{siam}
\bibliography{iscf,references-cut}

\begin{thebibliography}{10}

\bibitem{BollobasT98}
{\sc B.~Bollob{\'{a}}s and A.~Thomason}, {\em Proof of a conjecture of mader,
  erd{\"{o}}s and hajnal on topological complete subgraphs}, Eur. J. Comb., 19
  (1998), pp.~883--887.

\bibitem{BousquetDT11}
{\sc N.~Bousquet, J.~Daligault, and S.~Thomass{\'e}}, {\em Multicut is fpt}, in
  STOC, 2011, pp.~459--468.

\bibitem{ChenLL09}
{\sc J.~Chen, Y.~Liu, and S.~Lu}, {\em An improved parameterized algorithm for
  the minimum node multiway cut problem}, Algorithmica, 55 (2009), pp.~1--13.

\bibitem{chen2008fixed}
{\sc J.~Chen, Y.~Liu, S.~Lu, B.~O'sullivan, and I.~Razgon}, {\em A
  fixed-parameter algorithm for the directed feedback vertex set problem},
  Journal of the ACM (JACM), 55 (2008), p.~21.

\bibitem{Chitnis:2012DSFVS}
{\sc R.~H. Chitnis, M.~Cygan, M.~T. Hajiaghayi, and D.~Marx}, {\em Directed
  subset feedback vertex set is fixed-parameter tractable}, {ACM} Transactions
  on Algorithms, 11 (2015), p.~28.

\bibitem{ChitnisHM13}
{\sc R.~H. Chitnis, M.~Hajiaghayi, and D.~Marx}, {\em Fixed-parameter
  tractability of directed multiway cut parameterized by the size of the
  cutset}, {SIAM} J. Comput., 42 (2013), pp.~1674--1696.

\bibitem{ChoiNR89}
{\sc H.~Choi, K.~Nakajima, and C.~S. Rim}, {\em Graph bipartization and via
  minimization}, {SIAM} J. Discrete Math., 2 (1989), pp.~38--47.

\bibitem{CyganFKLMPPS15}
{\sc M.~Cygan, F.~V. Fomin, L.~Kowalik, D.~Lokshtanov, D.~Marx, M.~Pilipczuk,
  M.~Pilipczuk, and S.~Saurabh}, {\em Parameterized Algorithms}, Springer,
  2015.

\bibitem{CyganPPW13}
{\sc M.~Cygan, M.~Pilipczuk, M.~Pilipczuk, and J.~O. Wojtaszczyk}, {\em Subset
  feedback vertex set is fixed-parameter tractable}, SIAM J. Discrete Math., 27
  (2013), pp.~290--309.

\bibitem{edgemultiwaycuthardness}
{\sc E.~Dahlhaus, D.~S. Johnson, C.~H. Papadimitriou, P.~D. Seymour, and
  M.~Yannakakis}, {\em The complexity of multiterminal cuts}, Siam Journal on
  Computing, 23 (1994), pp.~864--894.

\bibitem{DemGMS07}
{\sc E.~D. Demaine, G.~Gutin, D.~Marx, and U.~Stege}, {\em 07281 open problems
  -- structure theory and {FPT} algorithmcs for graphs, digraphs and
  hypergraphs}, in Structure Theory and {FPT} Algorithmics for Graphs, Digraphs
  and Hypergraphs, 08.07. - 13.07.2007, 2007.

\bibitem{diestel}
{\sc R.~Diestel}, {\em Graph Theory}, Springer, Berlin, second ed.,
  electronic~ed., February 2000.

\bibitem{FominLPS16}
{\sc F.~V. Fomin, D.~Lokshtanov, F.~Panolan, and S.~Saurabh}, {\em Efficient
  computation of representative families with applications in parameterized and
  exact algorithms}, J. {ACM}, 63 (2016), pp.~29:1--29:60.

\bibitem{ford1956maximal}
{\sc L.~R. Ford and D.~R. Fulkerson}, {\em Maximal flow through a network},
  Canadian journal of Mathematics, 8 (1956), pp.~399--404.

\bibitem{FredmanKS84}
{\sc M.~L. Fredman, J.~Koml{\'{o}}s, and E.~Szemer{\'{e}}di}, {\em Storing a
  sparse table with 0(1) worst case access time}, J. {ACM}, 31 (1984),
  pp.~538--544.

\bibitem{GroheKS13}
{\sc M.~Grohe, S.~Kreutzer, and S.~Siebertz}, {\em Characterisations of nowhere
  dense graphs (invited talk)}, in {IARCS} Annual Conference on Foundations of
  Software Technology and Theoretical Computer Science, {FSTTCS} 2013,, vol.~24
  of LIPIcs, 2013, pp.~21--40.

\bibitem{KawarabayashiR10}
{\sc K.~ichi Kawarabayashi and B.~A. Reed}, {\em An (almost) linear time
  algorithm for odd cycles transversal}, in SODA, 2010, pp.~365--378.

\bibitem{IwataOY14}
{\sc Y.~Iwata, K.~Oka, and Y.~Yoshida}, {\em Linear-time {FPT} algorithms via
  network flow}, in SODA, 2014, pp.~1749--1761.

\bibitem{karp1972reducibility}
{\sc R.~M. Karp}, {\em Reducibility among combinatorial problems}, in
  Complexity of computer computations, Springer, 1972, pp.~85--103.

\bibitem{KomloS96}
{\sc J.~Koml{\'{o}}s and E.~Szemer{\'{e}}di}, {\em Topological cliques in
  graphs {II}}, Combinatorics, Probability {\&} Computing, 5 (1996),
  pp.~79--90.

\bibitem{Kratsch:2012MCDAG}
{\sc S.~Kratsch, M.~Pilipczuk, M.~Pilipczuk, and M.~Wahlstr{\"{o}}m}, {\em
  Fixed-parameter tractability of multicut in directed acyclic graphs}, {SIAM}
  J. Discrete Math., 29 (2015), pp.~122--144.

\bibitem{KratschW12}
{\sc S.~Kratsch and M.~Wahlstr{\"{o}}m}, {\em Representative sets and
  irrelevant vertices: New tools for kernelization}, in 53rd Annual {IEEE}
  Symposium on Foundations of Computer Science, {FOCS} 2012, New Brunswick, NJ,
  USA, October 20-23, 2012, {IEEE} Computer Society, 2012, pp.~450--459.

\bibitem{LokshtanovM13}
{\sc D.~Lokshtanov and D.~Marx}, {\em Clustering with local restrictions}, Inf.
  Comput., 222 (2013), pp.~278--292.

\bibitem{LokshtanovNRRS14}
{\sc D.~Lokshtanov, N.~S. Narayanaswamy, V.~Raman, M.~S. Ramanujan, and
  S.~Saurabh}, {\em Faster parameterized algorithms using linear programming},
  {ACM} Trans. Algorithms, 11 (2014), pp.~15:1--15:31.

\bibitem{LokshtanovR12}
{\sc D.~Lokshtanov and M.~S. Ramanujan}, {\em Parameterized tractability of
  multiway cut with parity constraints}, in Automata, Languages, and
  Programming - 39th International Colloquium, {ICALP} 2012, Warwick, UK, July
  9-13, 2012, Proceedings, Part {I}, 2012, pp.~750--761.

\bibitem{LokshtanovRS15}
{\sc D.~Lokshtanov, M.~S. Ramanujan, and S.~Saurabh}, {\em Linear time
  parameterized algorithms for subset feedback vertex set}, in Automata,
  Languages, and Programming - 42nd International Colloquium, {ICALP} 2015,
  Kyoto, Japan, July 6-10, 2015, Proceedings, Part {I}, 2015, pp.~935--946.

\bibitem{LokshtanovRS16}
\leavevmode\vrule height 2pt depth -1.6pt width 23pt, {\em A linear time
  parameterized algorithm for directed feedback vertex set}, CoRR,
  abs/1609.04347 (2016).

\bibitem{LokshtanovRSULC16}
\leavevmode\vrule height 2pt depth -1.6pt width 23pt, {\em A linear time
  parameterized algorithm for node unique label cover}, CoRR, abs/1604.08764
  (2016).

\bibitem{Mader1972}
{\sc W.~Mader}, {\em Existenzn-fach zusammenh{\"a}ngender teilgraphen in
  graphen gen{\"u}gend gro{\ss}er kantendichte}, Abhandlungen aus dem
  Mathematischen Seminar der Universit{\"a}t Hamburg, 37 (1972), pp.~86--97.

\bibitem{Marx06i}
{\sc D.~Marx}, {\em Parameterized graph separation problems}, Theor. Comput.
  Sci., 351 (2006), pp.~394--406.

\bibitem{marx2013finding}
{\sc D.~Marx, B.~O'sullivan, and I.~Razgon}, {\em Finding small separators in
  linear time via treewidth reduction}, ACM Transactions on Algorithms (TALG),
  9 (2013), p.~30.

\bibitem{MarxR14}
{\sc D.~Marx and I.~Razgon}, {\em Fixed-parameter tractability of multicut
  parameterized by the size of the cutset}, {SIAM} J. Comput., 43 (2014),
  pp.~355--388.

\bibitem{matula1983smallest}
{\sc D.~W. Matula and L.~L. Beck}, {\em Smallest-last ordering and clustering
  and graph coloring algorithms}, Journal of the ACM (JACM), 30 (1983),
  pp.~417--427.

\bibitem{misra2012parameterized}
{\sc N.~Misra, G.~Philip, V.~Raman, and S.~Saurabh}, {\em On parameterized
  independent feedback vertex set}, Theoretical Computer Science, 461 (2012),
  pp.~65--75.

\bibitem{nevsetvril2008grad}
{\sc J.~Ne{\v{s}}et{\v{r}}il and P.~O. de~Mendez}, {\em Grad and classes with
  bounded expansion i. decompositions}, European Journal of Combinatorics, 29
  (2008), pp.~760--776.

\bibitem{nevsetvril2011nowhere}
\leavevmode\vrule height 2pt depth -1.6pt width 23pt, {\em On nowhere dense
  graphs}, European Journal of Combinatorics, 32 (2011), pp.~600--617.

\bibitem{DBLP:books/daglib/0030491}
{\sc J.~Nesetril and P.~O. de~Mendez}, {\em Sparsity - Graphs, Structures, and
  Algorithms}, vol.~28 of Algorithms and combinatorics, Springer, 2012.

\bibitem{ramanujan2014linear}
{\sc M.~Ramanujan and S.~Saurabh}, {\em Linear time parameterized algorithms
  via skew-symmetric multicuts}, in Proceedings of the Twenty-Fifth Annual
  ACM-SIAM Symposium on Discrete Algorithms, Society for Industrial and Applied
  Mathematics, 2014, pp.~1739--1748.

\bibitem{ReedSV04}
{\sc B.~A. Reed, K.~Smith, and A.~Vetta}, {\em Finding odd cycle transversals},
  Oper. Res. Lett., 32 (2004), pp.~299--301.

\bibitem{RobertsonS95b}
{\sc N.~Robertson and P.~D. Seymour}, {\em Graph minors. {XIII.} {T}he disjoint
  paths problem}, J. Comb. Theory, Ser. B, 63 (1995), pp.~65--110.

\bibitem{stoer1997simple}
{\sc M.~Stoer and F.~Wagner}, {\em A simple min-cut algorithm}, Journal of the
  ACM (JACM), 44 (1997), pp.~585--591.

\bibitem{Sudakov16}
{\sc B.~Sudakov}, {\em Graph theory}, Lecture Notes,  (2016).

\end{thebibliography}

\end{document}